\documentclass[useAMS,usenatbib,usegraphicx]{mn2e}
\usepackage{times}
\def\kms{$\mbox{km s}^{-1}$}
\newcommand{\sauron}{{\texttt {SAURON}}}
\newcommand{\Xsauron}{{\texttt {XSAURON}}}
\newcommand{\oasis}{{\texttt {OASIS}}}
\newcounter{subfigure}

\title[The \sauron\ project -~III]
{The SAURON project - III. Integral-field absorption-line kinematics of 48 elliptical and lenticular galaxies}
\author[Emsellem et al.] {Eric\ Emsellem$^1$, Michele Cappellari$^2$, Reynier\ F.\ Peletier$^{1,3,4}$, Richard\  M.\ McDermid$^{2,5}$, 
\newauthor R.\ Bacon$^1$, M.\ Bureau$^{7}$\thanks{Hubble Fellow}, Y.\ Copin$^{1,6}$, Roger\ L.\ Davies$^{5,8}$,
Davor\ Krajnovi\'c$^2$,
\newauthor Harald\ Kuntschner$^9$, Bryan W. Miller$^{10}$, and P.\ Tim\ de Zeeuw$^2$ \\
$^1$Centre de Recherche Astronomique de Lyon, 9~Avenue Charles Andr\'e,
    69561 Saint Genis Laval, France\\
$^2$Sterrewacht Leiden, Niels Bohrweg~2, 2333~CA Leiden, The Netherlands\\
$^3$Department of Physics and Astronomy, University of Nottingham,
    University Park, Nottingham NG7~2RD, United Kingdom \\
$^4$Kapteyn Astronomical Institute, Postbus 800, 9700 AV Groningen, The Netherlands \\
$^5$Physics Department, University of Durham, South Road,
    Durham DH1~3LE, United Kingdom\\
$^6$Institut de Physique Nucl\'eaire de Lyon, 69622 Villeurbanne, France\\
$^7$Columbia Astrophysics Laboratory, 550~West 120th~Street, 1027 Pupin
    Hall, MC~5247, New York, NY~10027, USA\\
$^8$Denys Wilkinson Building, University of Oxford, Keble Road, Oxford,
     United Kingdom \\
$^9$Space Telescope European Coordinating Facility, European Southern Observatory, Karl-Schwarzschild-Str~2,
    85748 Garching, Germany\\
$^{10}$Gemini Observatory, Casilla 603, La Serena, Chile}
\pagerange{\pageref{firstpage}--\pageref{lastpage}} \pubyear{2004}

\begin{document}
\label{firstpage}
\maketitle
%
%
\begin{abstract}
We present the stellar kinematics of $48$ representative elliptical and lenticular galaxies obtained
with our custom-built integral-field spectrograph \sauron\ operating on the William Herschel
Telescope. The data were homogeneously processed through a dedicated reduction and analysis
pipeline. All resulting \sauron\ datacubes were spatially binned to a constant minimum
signal-to-noise. We have measured the stellar kinematics with an optimized (penalized pixel-fitting)
routine which fits the spectra in pixel space, via the use of optimal templates,
and prevents the presence of emission lines to
affect the measurements. We have thus generated maps of the mean stellar velocity $V$, the velocity dispersion
$\sigma$, and the Gauss--Hermite moments $h_3$ and $h_4$ of the line-of-sight velocity distributions.
The maps extend to approximately one effective radius. Many objects display kinematic twists,
kinematically decoupled components, central stellar disks, and other peculiarities, the nature of
which will be discussed in future papers of this series.
\end{abstract}
%
%
\section{INTRODUCTION}
\label{sec:intro}
We are carrying out a study of the structure of 72 representative nearby early-type galaxies and
spiral bulges based on measurements of the two-dimensional (2D) kinematics and line-strengths of stars
and gas with \sauron, a custom-built panoramic integral-field spectrograph for the William Herschel
Telescope (WHT), La Palma. \sauron\ is based on the {\texttt {TIGER}} microlens concept
(\citeauthor{betal95} \citeyear{betal95}; see also \citeauthor{becm00} \citeyear{becm00}), and is described in detail in Paper I
(\citeauthor*{betal01b} \citeyear{betal01b}). The objectives of the \sauron\ survey are summarized
in Paper II (\citeauthor*{zetal02} \citeyear{zetal02}), which also contains the definition and
global properties of the sample. Here we present maps of the stellar kinematics for the 48
elliptical (E) and lenticular (S0) galaxies in the survey.  The morphology and kinematics of the
ionized gas, the line-strength distributions, similar measurements for the spiral galaxies in the sample, and higher spatial resolution integral-field observations obtained with \oasis\ on the Canada-France-Hawaii Telescope (CFHT) will be presented separately. The data and maps presented in this paper will be made available via the \sauron\ WEB page http://www.strw.leidenuniv.nl/sauron/.

In Section~\ref{sec:observations} we summarize our observational campaign,
as well as the principles of the data reduction and the measurement of
the line-of-sight velocity distributions. We present the
kinematic maps in Section~\ref{sec:maps}, and our conclusions in
Section~\ref{sec:conclusions}.
\vfill

\section{Observations and data reduction}
\label{sec:observations}
\begin{table}
\centering
\caption{The \sauron\ observing runs.}
\label{tab:runs}
\begin{tabular}{ccc}
\hline
Run  &Dates  &Clear    \\
\hline
1 & 14--20/02/1999    &3.5/7       \\
2 & 08--14/10/1999    &4/8         \\
3 & 27/03--03/04/2000 &3.5/8       \\
4 & 01--04/09/2000    &3/3         \\
5 & 14--25/03/2001    &11.5/12     \\
6 & 14--17/01/2002    &3/4         \\
7 & 10--19/04/2002    &5/10        \\
8 & 24--27/04/2003     &2/4        \\
\hline
\end{tabular}
\end{table}

\subsection{Observing runs\label{sec:runs}}

We were allocated a total of 56 nights on the WHT, split into 8 runs over four years, to observe the
complete \sauron\ representative sample of 72 objects. This substantial allocation was divided
almost equally between the observing times of the United Kingdom and the Netherlands.  Table~\ref{tab:runs} summarizes the dates and
provides weather statistics. A total of 35.5 nights were clear. Run 1 followed soon after the
commissioning period (February 1--5, 1999, see \citeauthor{betal01b}). \sauron\ is a visitor
instrument, so each run was preceded by one or two days of configuring and calibrating the
instrument, with help from the Isaac
Newton Group staff. The first three runs were plagued by poor weather, and
technical problems further hindered progress in Run~4. 
The installation of {\tt ULTRADAS} (observing control software at the WHT) in the spring of 2001 improved the efficiency, which,
combined with the exceptional weather statistics of Run 5 allowed us to make more rapid
progress. A new volume-phase holographic grating (VPH) was installed for Run~8, resulting in a
further increase in the efficiency of the spectrograph to a remarkable 20\% end-to-end
(atmosphere, telescope, detector).

\begin{table*}
\caption{Details on the exposures of the E/S0 \sauron\ representative sample.
Characteristics for each target are provided in Paper II (\citeauthor*{zetal02} \citeyear{zetal02}).}
\label{tab:expo}
\begin{tabular}{lccc|lccc|lccc|lccc}
\hline
NGC &Run &~$\#$ &T$_{\rm exp}\hspace*{0.6cm}$ & NGC &Run &~$\#$ &T$_{\rm exp}\hspace*{0.6cm}$ & NGC &Run &~$\#$ &T$_{\rm exp}\hspace*{0.6cm}$ & NGC &Run &~$\#$ &T$_{\rm exp}$\\
~(1) &(2) &(3) &(4)&  ~(1) &(2) &(3) &(4) & ~(1) &(2) &(3) &(4)& ~(1) &(2) &(3) &(4)  \\
\hline

474             &2 &1 &4x1800 \hspace*{0.6cm}& 3379            &1 &1 &3x1800 \hspace*{0.6cm}& 4473            &5 &1 &4x1800 \hspace*{0.6cm}& 5813            &3 &1 &2x1800 \\
524             &4 &1 &2x1800 \hspace*{0.6cm}&                 &1 &2 &3x1800 \hspace*{0.6cm}&                 &5 &2 &4x1800 \hspace*{0.6cm}&                 &3 &2 &4x1800 \\
                &4 &2 &4x1800 \hspace*{0.6cm}&                 &1 &3 &2x1800 \hspace*{0.6cm}& 4477            &5 &1 &4x1800 \hspace*{0.6cm}&                 &3 &3 &4x1800 \\
                &4 &3 &4x1800 \hspace*{0.6cm}& 3384            &3 &1 &4x1800 \hspace*{0.6cm}&                 &5 &2 &4x1800 \hspace*{0.6cm}& 5831            &5 &1 &1x1800 \\
821             &2 &1 &3x1800 \hspace*{0.6cm}& 3414            &6 &1 &4x1800 \hspace*{0.6cm}& 4486            &5 &1 &4x1800 \hspace*{0.6cm}&                 &5 &1 &1x1200 \\
                &2 &2 &3x1800 \hspace*{0.6cm}& 3489            &5 &1 &4x1800 \hspace*{0.6cm}&                 &5 &2 &4x1800 \hspace*{0.6cm}&                 &5 &2 &1x1200 \\
1023            &2 &1 &4x1800 \hspace*{0.6cm}& 3608            &1 &1 &4x1800 \hspace*{0.6cm}&                 &5 &3 &3x1800 \hspace*{0.6cm}&                 &5 &2 &2x1800 \\
                &2 &2 &4x1800 \hspace*{0.6cm}& 4150            &3 &1 &4x1800 \hspace*{0.6cm}& 4526            &5 &1 &4x1800 \hspace*{0.6cm}&                 &5 &3 &1x1500 \\
2549            &1 &1 &4x1800 \hspace*{0.6cm}& 4262            &6 &1 &4x1800 \hspace*{0.6cm}&                 &5 &2 &4x1800 \hspace*{0.6cm}& 5838            &5 &1 &4x1800 \\
2685            &5 &1 &4x1800 \hspace*{0.6cm}& 4270            &6 &1 &4x1800 \hspace*{0.6cm}& 4546            &7 &1 &4x1800 \hspace*{0.6cm}& 5845            &5 &1 &4x1800 \\
                &5 &2 &4x1800 \hspace*{0.6cm}& 4278            &1 &1 &4x1800 \hspace*{0.6cm}& 4550            &1 &1 &4x1800 \hspace*{0.6cm}& 5846            &5 &1 &4x1800 \\
2695            &6 &1 &4x1800 \hspace*{0.6cm}&                 &1 &2 &4x1800 \hspace*{0.6cm}& 4552            &1 &1 &3x1800 \hspace*{0.6cm}&                 &5 &2 &4x1800 \\
2699            &5 &1 &4x1800 \hspace*{0.6cm}& 4374            &5 &1 &4x1800 \hspace*{0.6cm}& 4564            &3 &1 &4x1800 \hspace*{0.6cm}&                 &5 &3 &4x1800 \\
2768            &5 &1 &4x1800 \hspace*{0.6cm}&                 &5 &2 &4x1800 \hspace*{0.6cm}& 4570            &6 &1 &5x1800 \hspace*{0.6cm}& 5982            &5 &1 &4x1800 \\
                &5 &2 &4x1800 \hspace*{0.6cm}&                 &5 &3 &4x1800 \hspace*{0.6cm}& 4621            &5 &1 &4x1800 \hspace*{0.6cm}&                 &5 &2 &4x1800 \\
2974            &5 &1 &4x1800 \hspace*{0.6cm}& 4382            &5 &1 &4x1800 \hspace*{0.6cm}&                 &5 &2 &4x1800 \hspace*{0.6cm}& 7332            &2 &1 &4x1800 \\
                &5 &2 &4x1800 \hspace*{0.6cm}&                 &5 &2 &4x1800 \hspace*{0.6cm}& 4660            &5 &1 &4x1800 \hspace*{0.6cm}& 7457            &2 &1 &3x1800 \\
3032            &5 &1 &3x2100 \hspace*{0.6cm}& 4387            &5 &1 &4x1800 \hspace*{0.6cm}& 5198            &5 &1 &4x1800 \hspace*{0.6cm}&                 &2 &1 &1x1600 \\
3156            &6 &1 &4x1800 \hspace*{0.6cm}& 4458            &5 &1 &4x1800 \hspace*{0.6cm}& 5308            &5 &1 &4x1800 \hspace*{0.6cm}&                 &2 &1 &1x1030 \\
3377            &1 &1 &4x1800 \hspace*{0.6cm}& 4459            &5 &1 &4x1800 \hspace*{0.6cm}&                 &  &  &       \hspace*{0.6cm}&                 &2 &2 &4x1800 \\
                &  &  &       \hspace*{0.6cm}&                 &5 &2 &4x1800 \hspace*{0.6cm}&                 &  &  &       \hspace*{0.6cm}&                 &  &  &       \\
\hline
\end{tabular}
\\
Notes:
(1)~NGC number.
(2)~Run number (see Table \ref{tab:runs}).
(3)~Pointing number.
(4)~Exposure time, in sec.
\end{table*}

The \sauron\ sample of 48 E and S0 galaxies is representative of nearby bright early-type galaxies
($cz\leq3000$~\kms; $M_B\leq-18$~mag).  As described in \citeauthor{zetal02}, it contains $24$
galaxies in each of the E and S0 subclasses, equally divided between `cluster' and `field' objects
(the former defined as belonging to the Virgo cluster, the Coma~I cloud, and the Leo~I group, and
the latter being galaxies outside these clusters), uniformly covering the plane of ellipticity   
$\epsilon$ versus absolute blue magnitude $M_B$. Tables~A1 and A2 in \citeauthor{zetal02} 
provide basic information on the objects.

We used the low resolution mode of \sauron, giving a field of view of
$33\arcsec\times41\arcsec$, fully sampled by 1431 square lenses $0\farcs94\times0\farcs94$ in size, each of
which produces a spectrum. Another 146 lenses sample a small region about $1\farcm9$ from the field
center for simultaneous observation of the sky background. The wavelength range 4800--5380~\AA\ is covered at
$4.2$~\AA\ spectral resolution (FWHM, $\sigma_{\rm inst}=108$~\kms) with a sampling of $1.1$~\AA\ per
pixel. This range includes a number of
important stellar absorption lines (e.g.\ H$\beta$, Mg$\,b$, Fe) and potential emission lines (e.g.\
        H$\beta$, [{\sc O$\,$iii}], [{\sc N$\,$i}]).\looseness=-2

\begin{figure*}
\begin{center}
  \includegraphics[scale=0.84,trim=0cm 0cm 0cm 0cm]{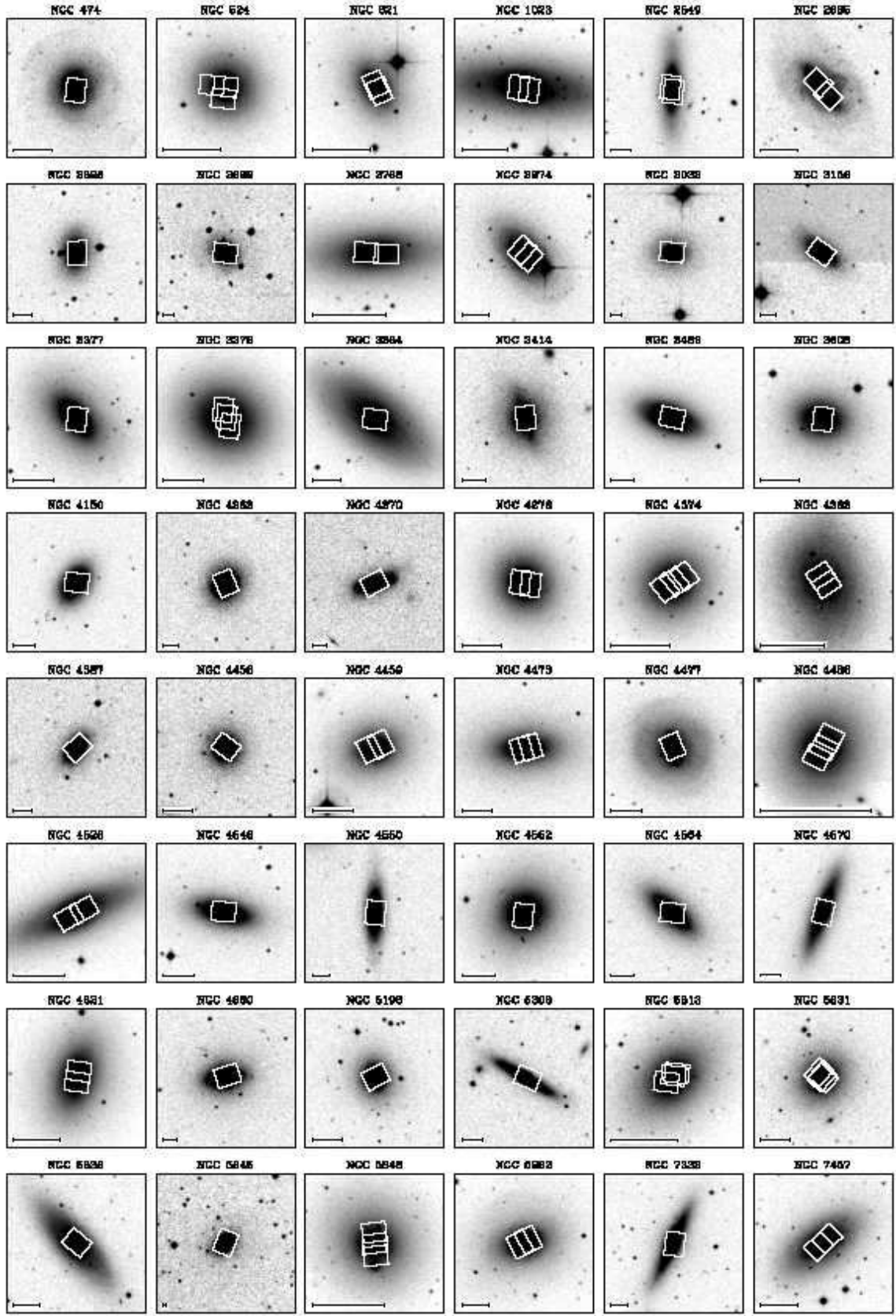}
\end{center}
\caption[]{$R$-band Digital Sky Survey images of all 48 E and S0 galaxies in the \sauron\
    representative sample. The size of each image is 4\arcmin$\times$4\arcmin, and the orientation is
    such that north is up and east is left. The bar located at the bottom-left corner of each
    image indicates the size of 1~$R_{\rm e}$ (from RC3: \citeauthor{RC3}~\citeyear{RC3})
    for that object where available. NGC~2699, NGC~4477 and NGC~5845 have $R_{\rm e}$ of 9\arcsec\
    \citep{Bla01}, 26\arcsec\ and 3\arcsec\ \citep{Alo03}, respectively.
    Overlaid on each image is the approximate footprint of the \sauron\ pointings obtained for that object.}
\label{fig:dss}
\end{figure*}

Each galaxy field was typically exposed for $4\times1800$~s, each dithered by a small
    non-integer number of lenses to avoid systematic errors due to e.g., bad CCD regions. In about
    one third of the cases, we
    constructed mosaics of two or three pointings to cover the galaxy out to about one effective
    radius $R_{\rm e}$, or, for the largest objects, out to $0.5R_{\rm e}$.  The footprints of these
    pointings are shown overlaid on $R$-band Digital Sky Survey images in Figure~\ref{fig:dss}.  For
    each of the 48 galaxies, Table~\ref{tab:expo} lists the run(s) in which it was observed, and the
    exposure times for the individual pointings. Note that Run~8 was used to complete the sample of
    early-type spirals, the data of which will be presented in a future paper of this series. Arc lamp 
	 exposures were taken before and after each 1800s exposure in order to track flexures and
    provide a wavelength calibration. We also obtained data
    for a number of velocity, line-strength and flux standard stars throughout each night, for calibration purposes.

\subsection{Data reduction}
\label{sec:reduction}

We reduced the \sauron\ observations with the dedicated \Xsauron\ software developed at CRAL and
described in \citeauthor{betal01b}. The steps include bias and dark subtraction, extraction of the
spectra using a fitted mask model, wavelength calibration, low frequency flat-fielding, cosmic-ray
removal, homogenization of the spectral resolution over the field, sky subtraction, and flux
calibration. The wavelength calibration is accurate to $0.1$~\AA\ ($6$~\kms).

\subsection{Merging and binning}
\label{sec:merging}

We merged multiple exposures and mosaiced different fields by
truncating the wavelength domain to a common range, registering the
exposures using reconstructed images, and combining the spectra (as
well as the noise spectra) with optimal weights and (re)normalisation.
During this process, we resampled the datacubes to a common spatial
scale of $0\farcs8\times0\farcs8$, reorienting them to
have the North up and East left, and correcting
for the effect of atmospheric refraction (as in \citeauthor{E+96}~\citeyear{E+96}).

Because a minimum signal-to-noise ratio ($S/N$) is required to reliably measure stellar kinematics,
we spatially binned the data cubes using an adaptive scheme developed by \citet{cc03}. In this
approach, the spectra are coadded by starting from the highest $S/N$ lenslet, and accreting
additional lenslets closest to the current bin centroid. A new bin is started each time the target
$S/N$ is reached. The resulting bin centroids are then used as starting points for a centroidal
Voronoi tessellation \citep[see, e.g.,][]{dfg99}, ensuring compact non-overlapping bins and a 
uniform $S/N$ in faint regions. We imposed a minimum $S/N$ of $60$ per spectral resolution element 
for the kinematical measurements presented here.

\subsection{Determination of the point spread function}
\label{sec:seeing}

We determined the point spread function (PSF) of each merged exposure by comparing the reconstructed
\sauron\ intensity distribution with HST/{\tt WFPC2} images. The \sauron\ PSF is
well-described by the sum of two circular Gaussians whose parameters are determined by minimizing
the difference between the {\tt WPFC2} image (convolved by these Gaussians) and the \sauron\
reconstructed image \citep{betal01a}. The result is valid in the central region where the overlap between
individual exposures is maximum. The derived seeing values are listed in Table~\ref{tab:allgal}.
{\tt WFPC2/F555W} images were favoured as the {\tt F555W} filter corresponds more closely to the \sauron\ spectral
domain. WFPC/F814W images were however used for about a third of the galaxies presented here,
for which the F555W filter was not available.
HST imaging does not exist for two objects in the E/S0 sample, namely NGC~2695 and NGC~4262, and
for another two, NGC\,4387 and NGC\,7332, only {\tt WFPC1} data is available. For these galaxies, we
favoured the estimates of the PSF width provided by the seeing monitor at WHT.
In most cases, \sauron\ undersamples the seeing.
\begin{table}
\caption{Characteristics of the E/S0 \sauron\ galaxies. }
\label{tab:allgal}
\begin{center}
\begin{tabular}{llrrlrr}
\hline
NGC &Type & V$_{\rm syst}$ & V$_{\rm bary}$ & Source & Seeing & $\theta_s$ \\
 ~(1) &~(2) &(3) &(4) &(5) &(6) &~(7)  \\
\hline
\noalign{\smallskip}
\phantom{0}474  &S0$^0$(s)       & 2319  &    0.9 & F814W & 1.9   &   0 \\
\phantom{0}524  &S0$^+$(rs)      & 2353  &  -30.0 & F555W & 1.4   &   0 \\
\phantom{0}821  &E6?             & 1722  &    7.6 & F555W & 1.7   &  30 \\
1023            &SB0$^-$(rs)     &  614  &   14.6 & F555W & 1.4   & 270 \\
2549            &S0$^0$(r)sp     & 1069  &  -12.5 & F702W & 1.7   &   0 \\
2685            &(R)SB0$^+$pec   &  883  &  -20.8 & F555W & 1.3   & 322 \\
2695            &SAB0$^0$(s)     & 1831  &   10.9 & WHT   & 2.0   &   9 \\
2699            &E:              & 1864  &  -17.3 & F702W & 1.3   & 270 \\
2768            &E6:             & 1359  &  -18.9 & F555W & 1.4   &   5 \\
2974            &E4              & 1886  &  -14.3 & F555W & 1.4   & 326 \\
3032            &SAB0$^0$(r)     & 1559  &  -20.6 & F606W & 1.5   & 270 \\
3156            &S0:             & 1541  &   18.8 & F814W & 1.6   &  61 \\
3377            &E5-6            &  690  &    5.0 & F555W & 2.1   &   0 \\
3379            &E1              &  916  &    4.8 & F555W & 1.8   & 270 \\
3384            &SB0$^-$(s):     &  737  &   -17.6 & F555W & 1.8  & 270 \\
3414            &S0~pec          & 1472  &   16.9 & F814W & 1.4   & 193 \\
3489            &SAB0$^+$(rs)    &  702  &   -8.5 & F814W & 1.1   &  84 \\
3608            &E2              & 1228  &    5.6 & F555W & 1.5   &   0 \\
4150            &S0$^0$(r)?      &  219  &   -8.9 & F814W & 2.1   & 270 \\
4262            &SB0$^-$(s)      & 1371  &   25.7 & WHT   & 2.6   &  30 \\
4270            &S0              & 2333  &   27.3 & F606W & 2.0   & 124 \\
4278            &E1-2            &  631  &    9.7 & F555W & 1.9   &   0 \\
4374            &E1              & 1023  &    2.8 & F814W & 2.2   &  45 \\
4382            &S0$^+$(s)pec    &  742  &   -1.8 & F555W & 2.7   & 309 \\
4387            &E               &  568  &    3.4 & WHT   & 1.3   & 320 \\
4458            &E0-1            &  683  &    2.3 & F555W & 1.6   &  60 \\
4459            &S0$^+$(r)       & 1200  &   -1.8 & F814W & 1.5   &  34 \\
4473            &E5              & 2249  &    0.2 & F555W & 1.9   &  24 \\
4477            &SB0(s):?        & 1350  &   -2.1 & F606W & 2.1   &  39 \\
4486            &E0-1$^+$pec     & 1274  &    2.5 & F555W & 1.0   &  70 \\
4526            &SAB0$^0$(s)     &  626  &    0.8 & F555W & 2.8   &  37 \\
4546            &SB0$^-$(s):     & 1051  &   -8.7 & F606W & 1.8   & 274 \\
4550            &SB0$^0$:sp      &  413  &  -28.1 & F555W & 2.1   &   0 \\
4552            &E0-1            &  351  &   16.2 & F555W & 1.9   &   0 \\
4564            &E               & 1149  &   -2.6 & F702W & 2.1   & 270 \\
4570            &S0~sp           & 1780  &   28.1 & F555W & 1.7   & 353 \\
4621            &E5              &  456  &    3.5 & F555W & 1.6   &  89 \\
4660            &E               & 1089  &    3.4 & F555W & 1.6   & 294 \\
5198            &E1-2:           & 2531  &   -1.1 & F702W & 1.5   & 304 \\
5308            &S0$^-$~sp       & 1985  &   -5.0 & F555W & 2.4   &  74 \\
5813            &E1-2            & 1947  &   15.6 & F555W & 1.7   & 270 \\
5831            &E3              & 1656  &   21.1 & F702W & 1.6   & 325 \\
5838            &S0$^-$          & 1340  &   20.7 & F814W & 1.6   &  57 \\
5845            &E:              & 1474  &   20.5 & F555W & 1.5   & 344 \\
5846            &E0-1            & 1710  &   19.6 & F555W & 1.4   & 284 \\
5982            &E3              & 2935  &    1.4 & F555W & 1.2   &  34 \\
7332            &S0~pec~sp       & 1206  &  -13.4 & WHT   & 1.1   &   0 \\
7457            &S0$^-$(rs)?     &  845  &   -9.0 & F555W & 1.3   & 320 \\

\hline
\end{tabular}
\end{center}
Notes:
(1)~NGC number.
(2)~Hubble type (RC3: \citeauthor{RC3} \citeyear{RC3})
(3)~Estimate of the heliocentric systemic velocity in \kms\ (see Sect.~\ref{sec:maps}).
(4)~Mean barycentric correction applied to the merged datacubes (in \kms).
(5)~Source for the seeing determination (see text) \
(6)~Seeing, full width at half maximum in arcsec.
(7)~Position angle, in degrees, of the vertical (upward) axis in the maps shown in
Figs.~\ref{fig:maps1}-\ref{fig:maps12}.
\end{table}

\subsection{Stellar kinematics}
\label{sec:kinematics}

Most of the galaxies in this sample are contaminated by significant emission lines  of e.g.,
H$\beta$, the [{\sc O$\,$iii}]$\lambda\lambda$4959,5007 or [{\sc N$\,$i}]$\lambda\lambda$5198,5200 doublets. This causes problems
for techniques working in Fourier space, e.g. the Fourier Correlation Quotient (FCQ,
\citeauthor*{b90} \citeyear{b90}) or the Cross Correlation Fitting (CCF,
\citeauthor{stat95} \citeyear {stat95}) techniques, as they do not permit direct masking
of specific spectral regions. Such methods can obviously make use of
continuous spectral regions not or weakly affected by emission lines. 
However, considering the short spectral domain
provided by \sauron, and the significant fraction of galaxies found
to exhibit the emission lines which punctuate this region, these
techniques are not well suited to these data.
For this reason we opted for the use of a direct pixel-fitting (PXF)
routine to extract the kinematics, allowing emission-line regions to be easily masked
\cite[e.g.,][and references therein]{vdM94,ce04}. We emphasize that all results obtained with
this new implementation are fully consistent with the early results presented in Paper~II.

\subsubsection{Penalized pixel fitting with optimal templates}
\label{sec:pxf}

We derive the line-of-sight velocity distribution (hereafter LOSVD) parametrized by a Gauss-Hermite
function \citep{vdM93,g93} for each spectrum using the PXF routine: the algorithm finds the best fit
to a galaxy spectrum (rebinned in $\ln\lambda$) by convolving a template stellar spectrum with the
corresponding LOSVD. This provides the mean velocity $V$ and the velocity dispersion $\sigma$, as
well as the higher order Gauss-Hermite moments $h_3$ and $h_4$, which quantify the asymmetric and
symmetric departures of the LOSVD from a pure Gaussian (related to the skewness and kurtosis
respectively). Although this appears as a conceptually simple procedure, there are two
critical issues (not specific to \sauron) to carefully consider to obtain robust and reliable
kinematic measurements.

\begin{description}
\item[\bf Optimal templates:] Although robust against the presence of emission
lines, PXF is sensitive to template mismatch, in contrast to CCF and FCQ
which minimize the coupling between different absorption features by restricting
the analysis to the main peak of the cross correlation between the template and
the galaxy spectra. The use of PXF thus requires templates which match closely
the galaxy spectrum under scrutiny. The derivation of optimal stellar templates
must be conducted for every individual spectrum (as galaxies may e.g. exhibit
		significant metallicity or age gradients within the \sauron\ field of
		view). This is achieved via the use of an extensive stellar library
spanning a large range of metallicities and ages. We primarily included 19
spectra from the library of single-metallicity stellar population models of
\citet{vaz99}, and added 5 spectra from the Jones stellar library (from which
		Vazdekis' models are built; \citeauthor{J97} \citeyear{J97}) to provide
spectra with strong Mg\,$b$ indices. This also requires the inclusion of
additive Legendre polynomials to adapt the continuum shape of the templates.

\item[\bf Penalized pixel fitting:] When measuring the Gauss-Hermite moments of
the LOSVD up to $h_4$, undersampling becomes important when the observed
dispersion is less than about 2 pixels (equivalent to around 120~\kms for
\sauron). Below this dispersion a very high
$S/N$ is required to accurately measure the Gauss-Hermite parameters. At our
minimum $S/N\approx60$, the data are therefore unable to significantly constrain all the
$(V,\sigma,h_3,h_4)$ parameters simultaneously in the undersampled regime, and the scatter in the
measurements increases dramatically. In this situation one usually wants to
reduce the parametric form of the LOSVD to a simple Gaussian. We therefore used
the penalized pixel-fitting (hereafter pPXF) method developed by \citet{ce04} to
perform this biasing of the solution towards a Gaussian in a statistically
motivated and automatic way, as a function of both the $S/N$ of the spectra, and
the observed $\sigma$.  \end{description}

We implemented both the template optimization and the penalization directly into
our PXF routine.  This works for each individual galaxy spectrum as follows. The
direct pixel fitting is a non-linear least-squares problem, for the parameters
$(V,\sigma,h_3,h_4)$ of the LOSVD, which is solved using a sequential quadratic
programming method including bounds (NAG routine E04UNF). At each iteration we
find the best fitting linear combination of the stellar templates (with positive
		weights) convolved with the corresponding LOSVD. This includes the
additive polynomials and provides the optimal template for that specific set of
kinematic parameters. The residual vector $\mathbf{r}$ between the observed
spectrum and the convolved optimal spectrum is then derived and the penalization
term is added to obtain a `perturbed' vector of residuals $\mathbf{r'}$, which is
fed into the nonlinear optimization routine.

The perturbed vector has the form
$\mathbf{r'}=\mathbf{r}+\lambda\mathcal{D}\sigma(\mathbf{r})$, where $\lambda$
is an adjustable parameter of order unity, $\mathcal{D}^2 = h_3^2 + h_4^2$ is
the integrated squared deviation of the LOSVD from its best fitting Gaussian
\citep{vdM93}, and $\sigma(\mathbf{r})$ is a robust estimate of the standard
deviation of the residuals. The qualitative interpretation of the perturbation
can be illustrated by noting that for $\lambda=1$, an amplitude of 0.1 for $h_3$
or $h_4$ (10\% deviation from a Gaussian) requires an improvement of at least
0.5\% of the unperturbed $\sigma(\mathbf{r})$ to be accepted by the optimization
routine as a decrease of the perturbed residuals $\mathbf{r'}$. Setting
$\lambda=0$ provides a non-penalized PXF routine. After extensive testing we
adopted a value of $\lambda =
0.7$, which makes sure the bias is always small compared to the measurement
  errors, when $\sigma\ga120$ \kms and $S/N\ga60$. See \citet{ce04} for details
  on this pPXF technique.

\begin{figure} 
\begin{center}
\includegraphics[width=\columnwidth]{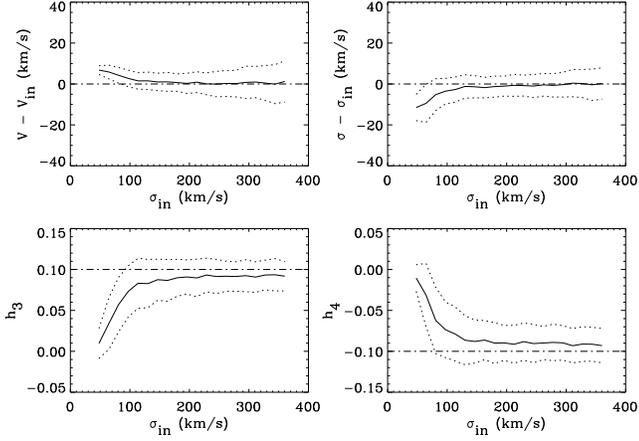} 
\end{center} 
\caption[]{Simulated errors on the measured kinematic parameters using Monte Carlo realizations.
These plots illustrate the case of an input $S/N$ of 60, $h_3 = 0.1$,
$h_4 = -0.1$, and for velocity dispersions between 50 and 360~\kms.  The
solid line in the top panels show the difference between the measured
(via pPXF) and the input values for the mean velocity $V$ (left) and
the velocity dispersion $\sigma$ (right).  In the bottom panels, we
show the recovered $h_3$ (left) and $h_4$ (right) values.  The dotted
lines illustrate the $1\sigma$ error on these measurements at constant
input velocity dispersion.}
\label{fig:errors}
\end{figure}

We estimated errors by a Monte-Carlo method in which the kinematic parameters
are derived from many realizations of the input spectrum obtained by adding
Poissonian noise to a model galaxy spectrum. Figure~\ref{fig:errors} summarizes
the results from such simulations. The significant damping of $h_3$ and $h_4$
towards zero is visible when $\sigma\la120$~\kms, but as noted above, this
damping effect decreases with increasing $S/N$ and becomes almost negligible for
$S/N \ga 100$, as we have in the center of most of our galaxies. For a $S/N$ of
60, the $1\sigma$ errors on $V$, $\sigma$, $h_3$ and $h_4$ are 4~\kms, 5~\kms,
0.03 and 0.03 respectively for $\sigma = 120$~\kms, and 8~\kms, 7~\kms, 0.02
and 0.02 respectively for $\sigma = 300$~\kms. At higher $S/N$, the errors
scale roughly inversely proportional to $S/N$. These estimates are only lower
limits since they do not account for the effect of template and continuum
mismatch (Appendix~\ref{sec:mismatch}). Error maps will be made available
with the public data release.  

\subsubsection{The robustness of the kinematic measurements}

In order to assess the robustness of the method described above, we conducted an extensive
series of tests. We first carefully inspected the fits of the observed 
spectra obtained for a number of galaxies and these were always found to be excellent 
(see Appendix~\ref{sec:fits} for some illustrations).
We checked that the kinematic measurements were not affected by the
details of the input library.

Most galaxies required a Legendre polynomial of degree 6 for the pixel fitting. For a few galaxies
we had to impose a higher degree (10) to properly fit the spectral shape. We verified that this did
not artificially bias the kinematic measurements, even the more sensitive higher order Gauss--Hermite
moments $h_3$ and $h_4$. We also found that applying either multiplicative or additive polynomials
led to perfectly consistent results.

All tests were performed with two completely independent implementations of the pPXF method and
both gave indistinguishable results. The stellar kinematics of all 48 galaxies was also extracted
using FCQ in combination with the gas-cleaning procedure described in Paper~II. For galaxies not
severely affected by gas emission the two methods produced fully consistent results (see
Section~\ref{sec:fcq}).

\subsubsection{Comparison with published kinematics}

\begin{figure}
\begin{center}
  \includegraphics[width=\columnwidth]{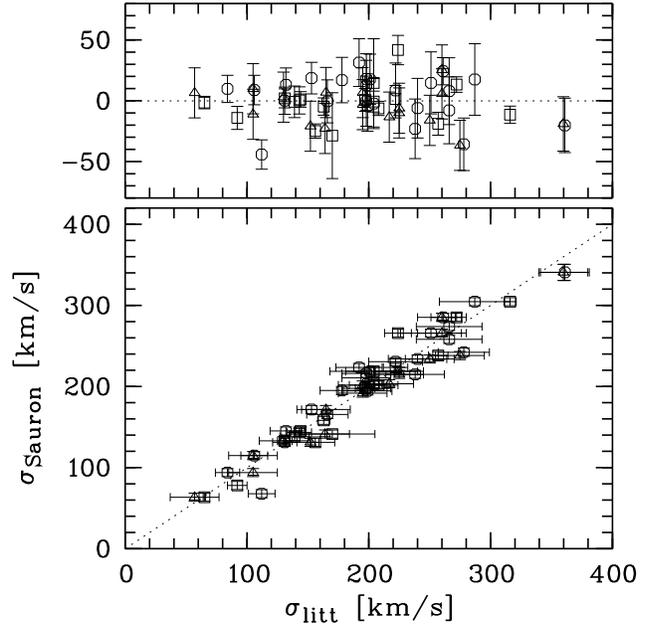}
\end{center}
\caption[]{Comparison between aperture ($R_e / 10$) measurements of the central dispersion of the \sauron\ E/S0
galaxies and values taken from the literature. {\em Bottom}: \sauron\ central dispersions versus
    values from \citet{7sam} (circles), \citet{Fab97} (triangles) and \citet{PS96} (squares).
{\em Top}: difference between the values from \sauron\ and the literature. The errors from both
datasets have been summed quadratically.}
\label{fig:sigma}
\end{figure}

We have already shown in Paper~II that the \sauron\ stellar and gas kinematics are 
consistent with published data.
To have a more global view of the sample, we performed a comparison between the central stellar
velocity dispersion $\sigma_0$ from the \sauron\ data and from published values. We made use of three
main sources which include a significant overlap with the \sauron\ target list, namely
the 7 Samurai (\citeauthor{7sam} \citeyear{7sam}), the weighted averaged
values compiled by \citet{Fab97} and the
compilation of data gathered from papers by Prugniel and Simien and available via
the Hypercat\footnote{http://www-obs.univ-lyon1.fr/hypercat/} on-line catalog
(\citeauthor{PS96} \citeyear{PS96}).
Our measurements were derived by averaging (luminosity-weighted) dispersion
values within a circular aperture with a radius of $R_e / 10$, but imposing
a minimum aperture of $2.4\arcsec \times 2.4\arcsec$ ($3\times3$ pixels).
A linear regression applied to the combined dataset provides a slope of $0.98 \pm 0.03$
and an average residual of $-1 \pm 17$~\kms. The average residuals for each individual
dataset are all consistent with zero and with the scatter expected from the quadratically summed
error bars (see Fig.~\ref{fig:sigma}): $3.0 \pm 18$~\kms\ for the 7 Samurai,
$-2.1 \pm 16$~\kms\ for \citet{Fab97} and $-5.6 \pm 15$~\kms\ for the data of \citet{PS96}.
There is no measurable trend for sub-samples defined either for $\sigma < 200$~\kms\ or
$\sigma > 200$~\kms. This comparison demonstrates that our (velocity dispersion) measurements
do not seem to suffer from any detectable systematics.
%
%
\renewcommand{\thefigure}{\arabic{figure}\alph{subfigure}}
\setcounter{subfigure}{1}

\begin{figure*}
\begin{center}
  \includegraphics[width=\textwidth,trim=0cm 0.5cm 0cm 0cm]{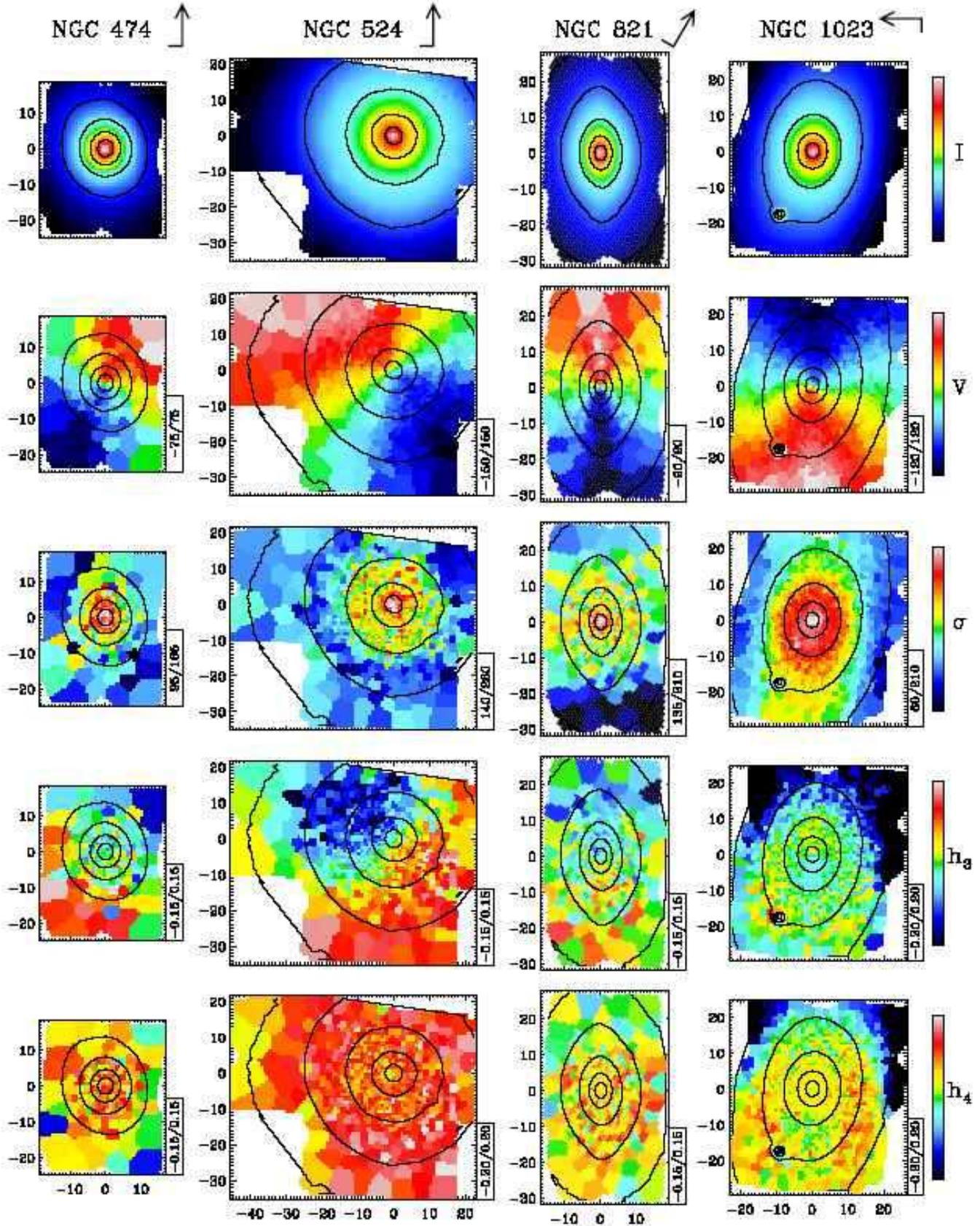}
\end{center}
\caption[]{Maps of the stellar kinematics of the 48 E and S0 galaxies in the \sauron\ representative
sample. The \sauron\ spectra have been spatially binned to a minimum $S/N$ of 60 by
means of the Voronoi 2D-binning algorithm of \protect\citet{cc03}. All maps are plotted to the
same spatial scale. The arrow and its associated dash at the top of each column mark the North
and East directions, respectively; the corresponding position angle of the vertical (upward)
axis is provided in Table~\ref{tab:allgal}. From top to bottom: i) reconstructed total
intensity, ii) stellar mean velocity $V$, iii) stellar velocity dispersion $\sigma$, iv) and
v) Gauss--Hermite moments $h_3$ and $h_4$. The cuts levels are indicated in a box on the right
hand side of each map. } 
\label{fig:maps1}
\end{figure*}

\addtocounter{figure}{-1}
\addtocounter{subfigure}{1}

\begin{figure*}
\begin{center}
  \includegraphics[width=\textwidth,trim=0cm 0cm 0cm 0cm]{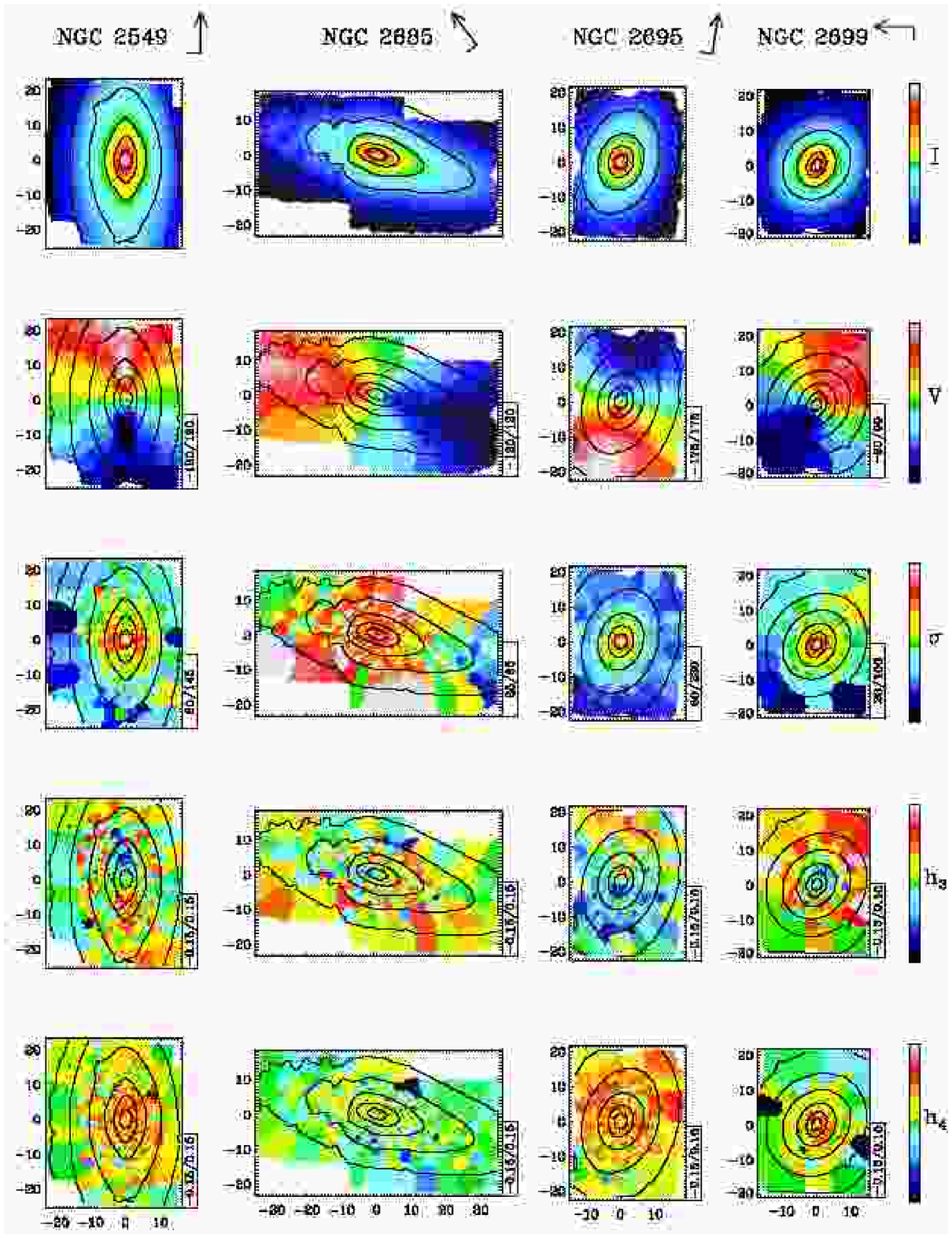}
\end{center}
\caption[]{}
\label{fig:maps2}
\end{figure*}

\addtocounter{figure}{-1}
\addtocounter{subfigure}{1}

\begin{figure*}
\begin{center}
  \includegraphics[width=\textwidth,trim=0cm 0cm 0cm 0cm]{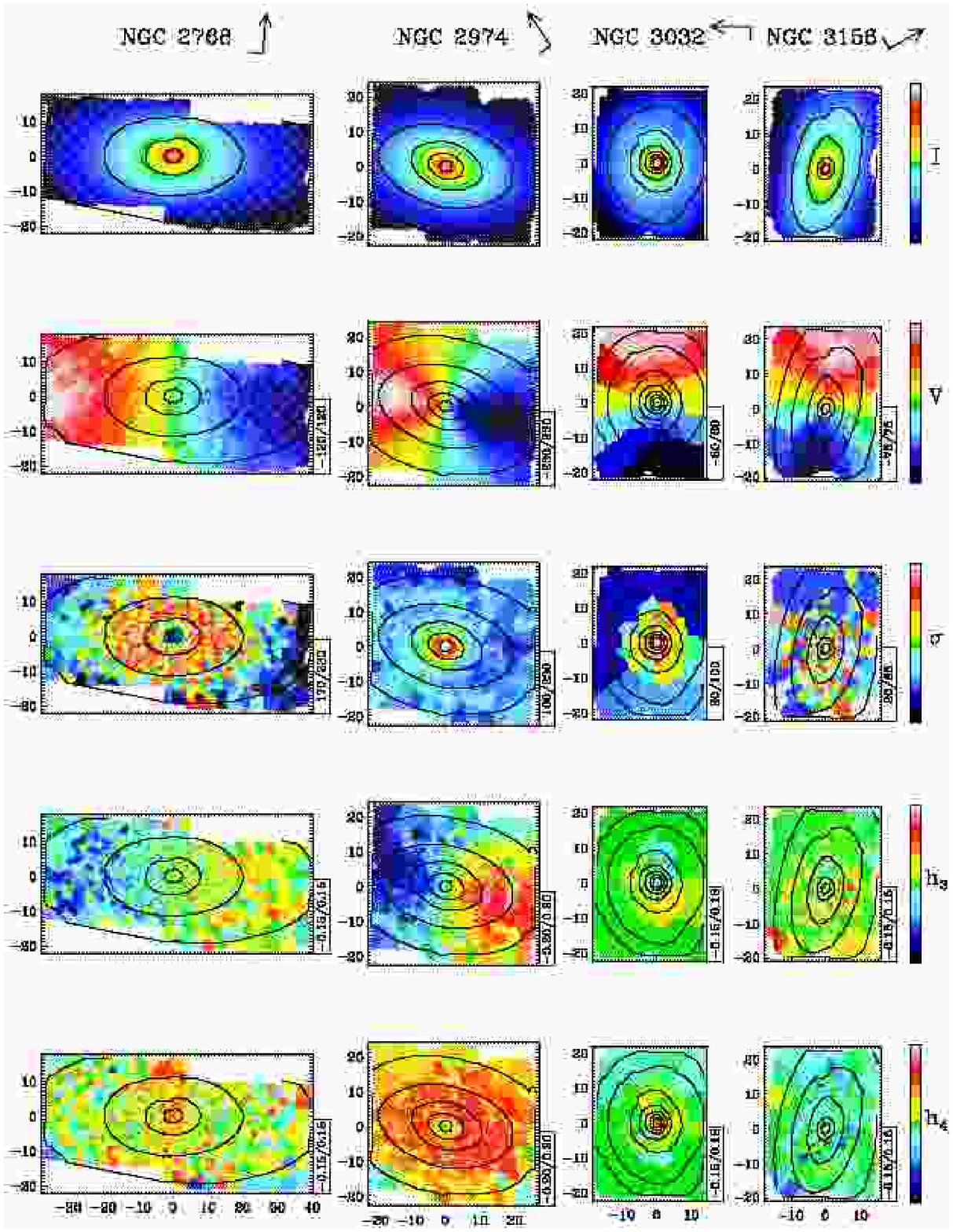}
\end{center}
\caption[]{}
\label{fig:maps3}
\end{figure*}

\addtocounter{figure}{-1}
\addtocounter{subfigure}{1}

\begin{figure*}
\begin{center}
  \includegraphics[width=\textwidth,trim=0cm 0cm 0cm 0cm]{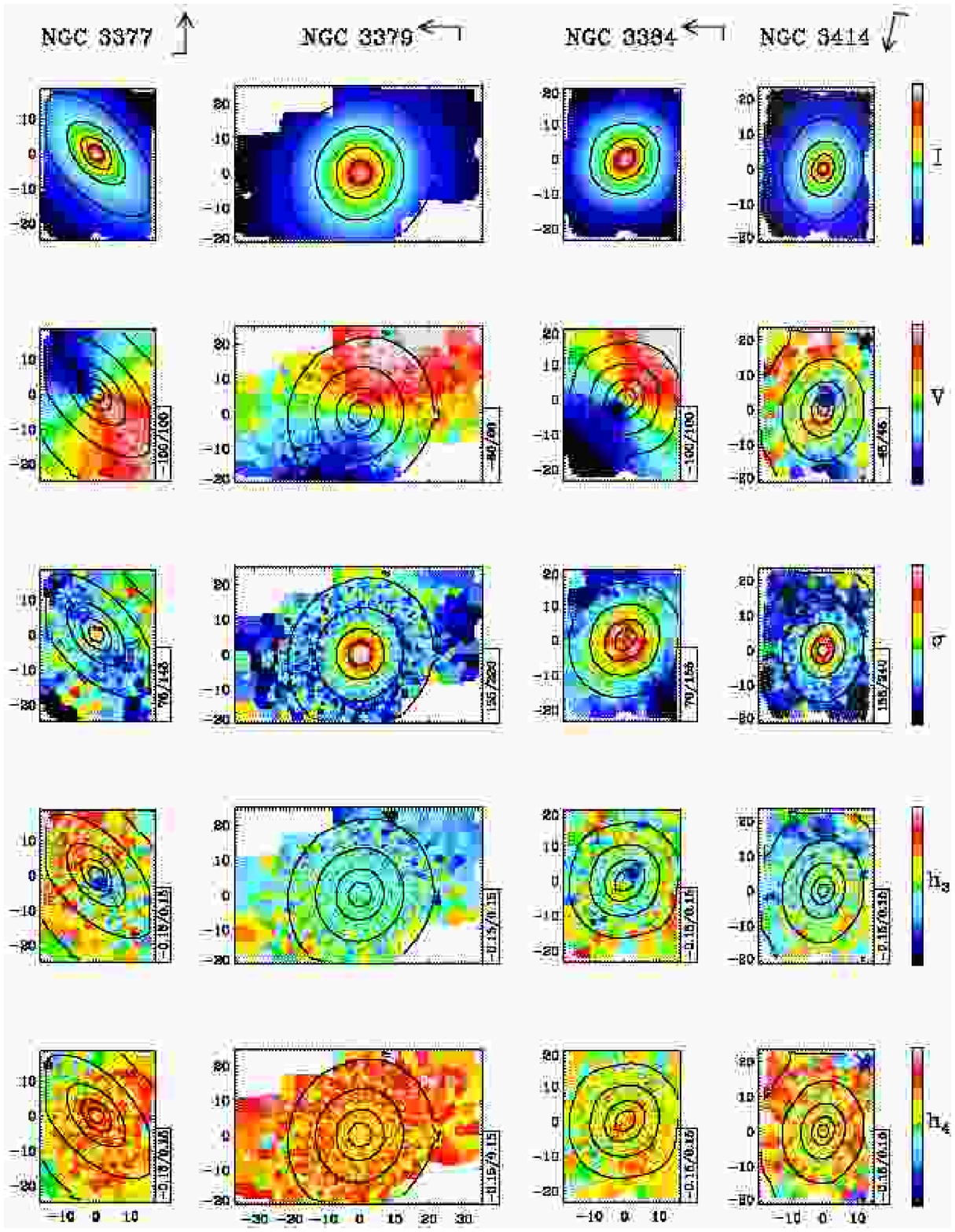}
\end{center}
\caption[]{}
\label{fig:maps4}
\end{figure*}

\addtocounter{figure}{-1}
\addtocounter{subfigure}{1}

\begin{figure*}
\begin{center}
  \includegraphics[width=\textwidth,trim=0cm 0cm 0cm 0cm]{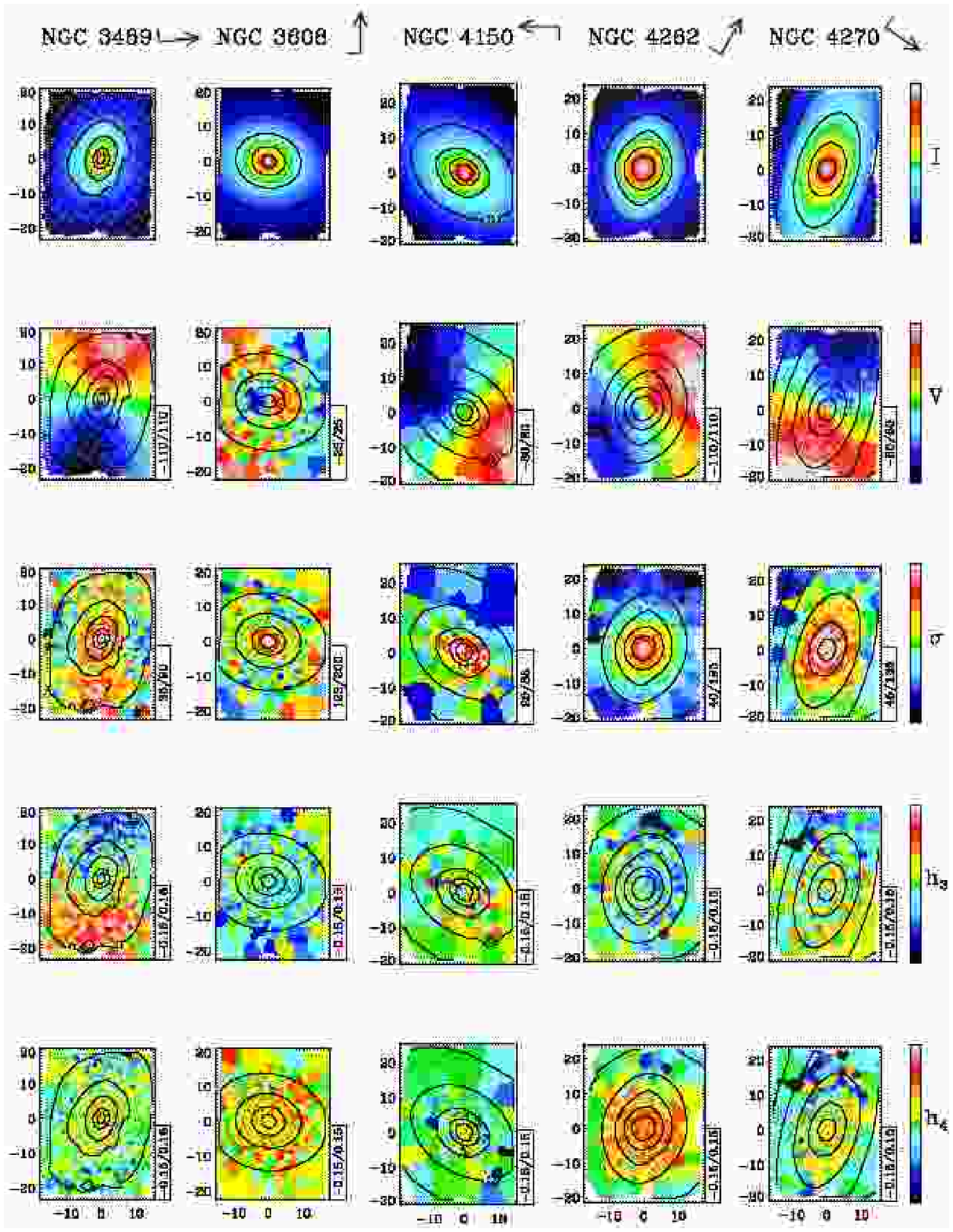}
\end{center}
\caption[]{}
\label{fig:maps5}
\end{figure*}

\addtocounter{figure}{-1}
\addtocounter{subfigure}{1}

\begin{figure*}
\begin{center}
  \includegraphics[width=\textwidth,trim=0cm 0cm 0cm 0cm]{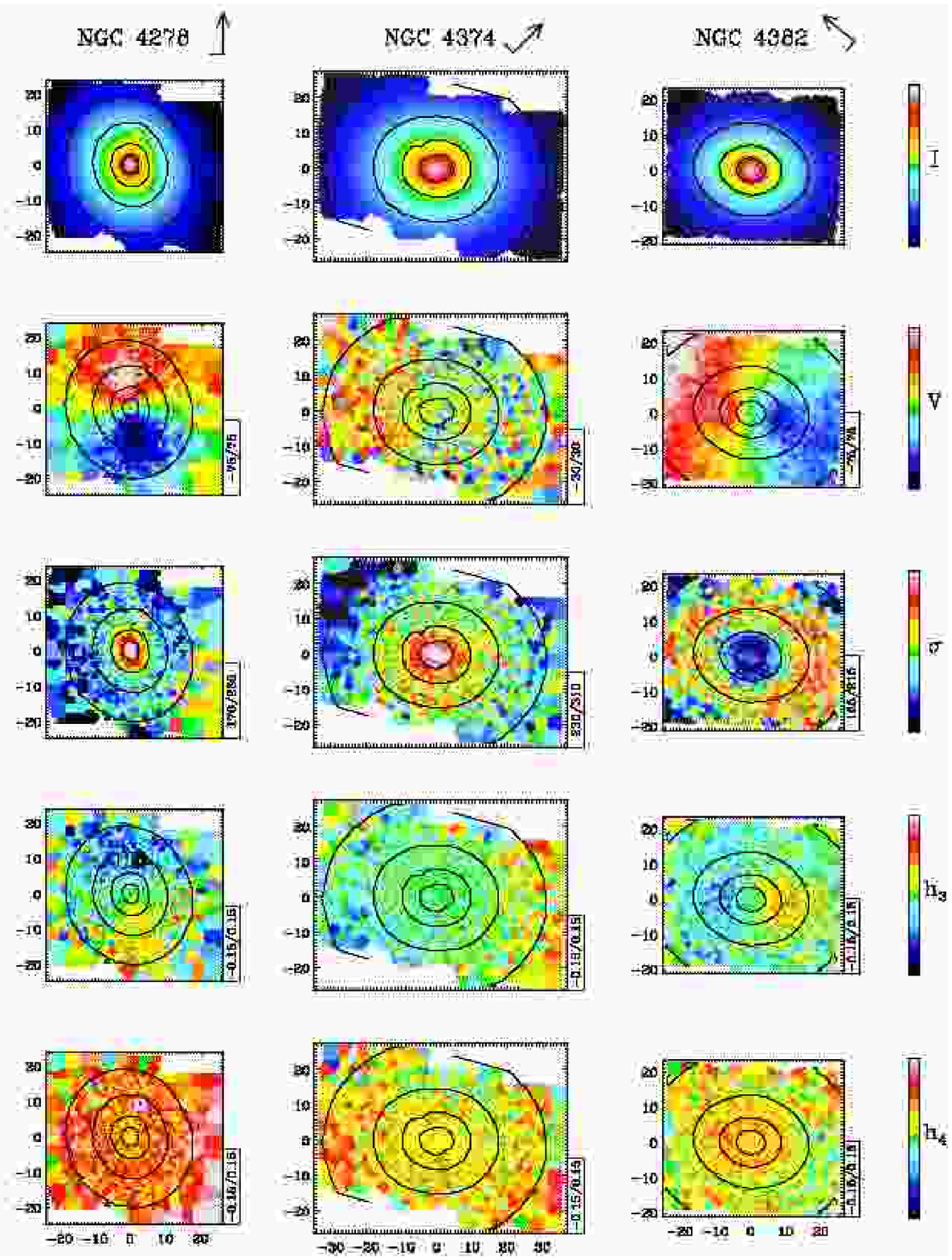}
\end{center}
\caption[]{}
\label{fig:maps6}
\end{figure*}

\addtocounter{figure}{-1}
\addtocounter{subfigure}{1}

\begin{figure*}
\begin{center}
  \includegraphics[width=\textwidth,trim=0cm 0cm 0cm 0cm]{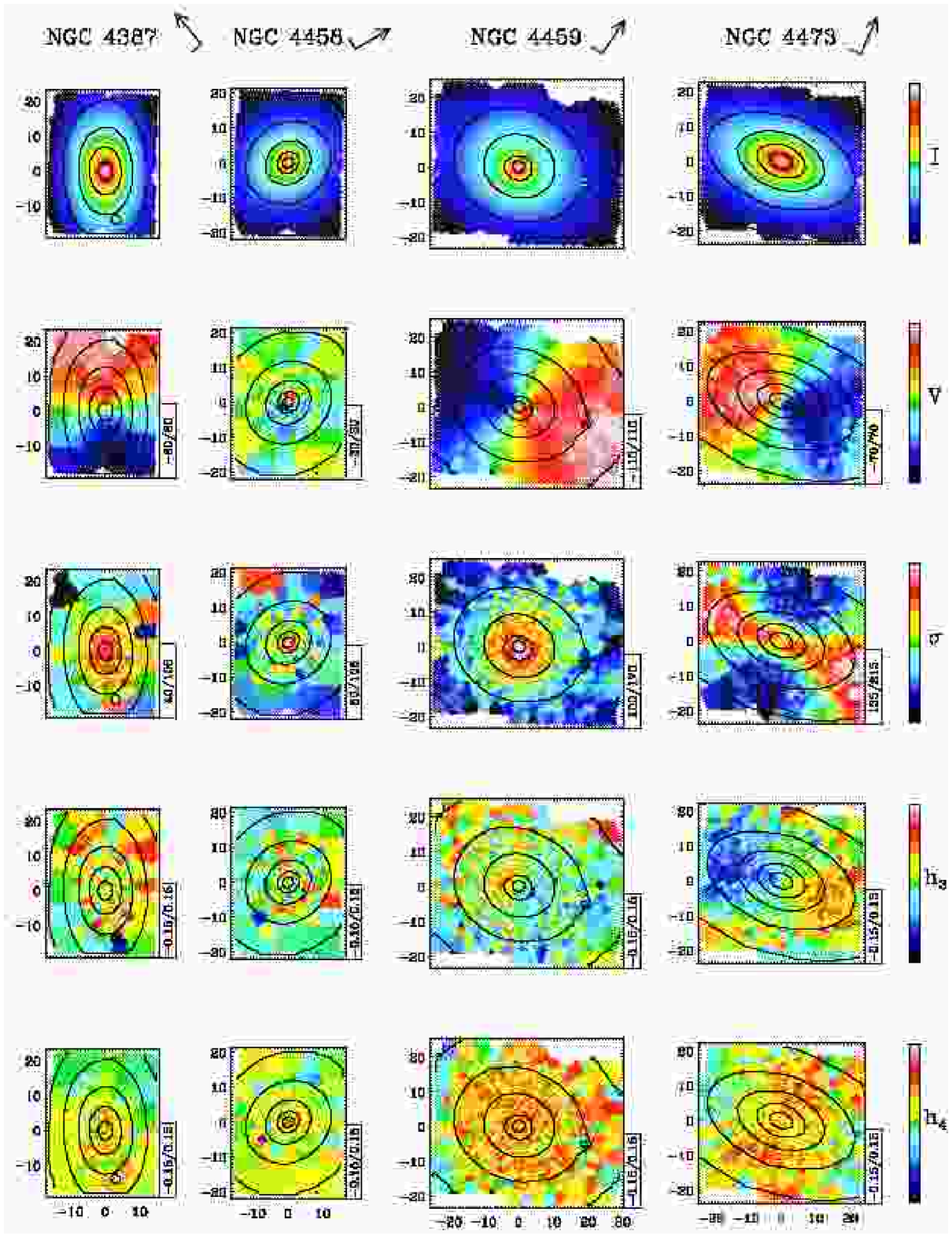}
\end{center}
\caption[]{}
\label{fig:maps7}
\end{figure*}

\addtocounter{figure}{-1}
\addtocounter{subfigure}{1}

\begin{figure*}
\begin{center}
  \includegraphics[width=\textwidth,trim=0cm 0cm 0cm 0cm]{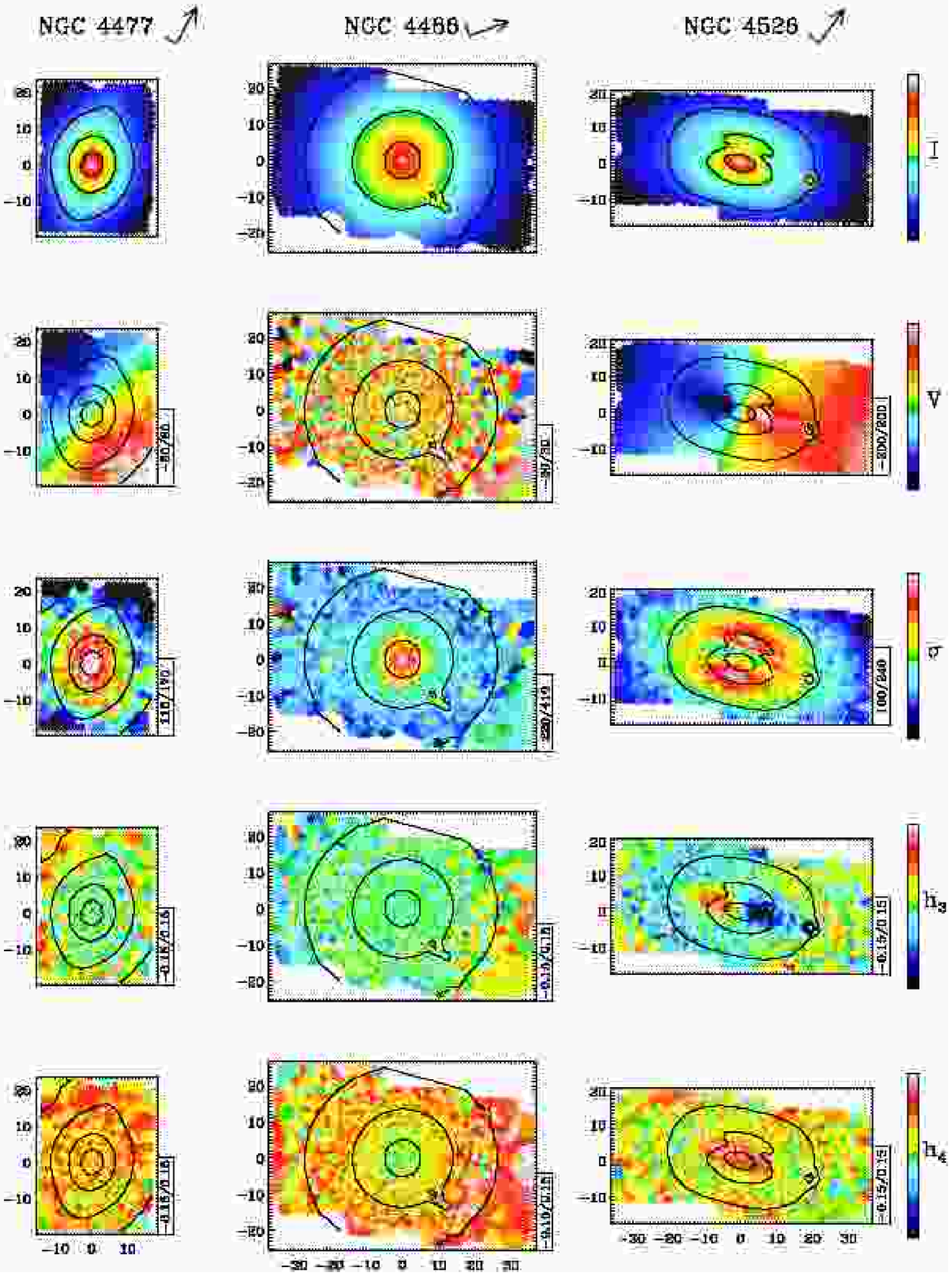}
\end{center}
\caption[]{}
\label{fig:maps8}
\end{figure*}

\addtocounter{figure}{-1}
\addtocounter{subfigure}{1}

\begin{figure*}
\begin{center}
  \includegraphics[width=\textwidth,trim=0cm 0cm 0cm 0cm]{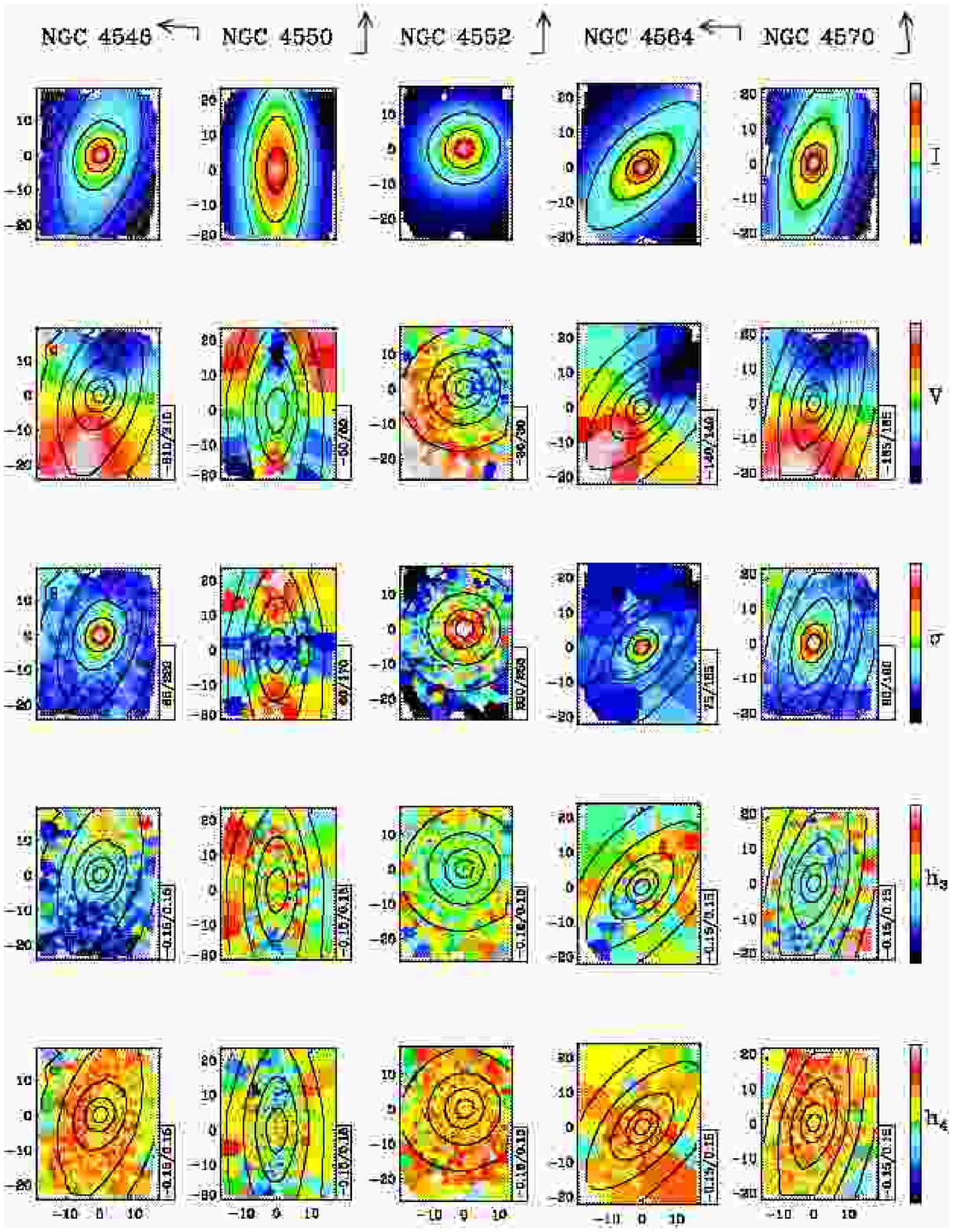}
\end{center}
\caption[]{}
\label{fig:maps9}
\end{figure*}

\addtocounter{figure}{-1}
\addtocounter{subfigure}{1}

\begin{figure*}
\begin{center}
  \includegraphics[width=\textwidth,trim=0cm 0cm 0cm 0cm]{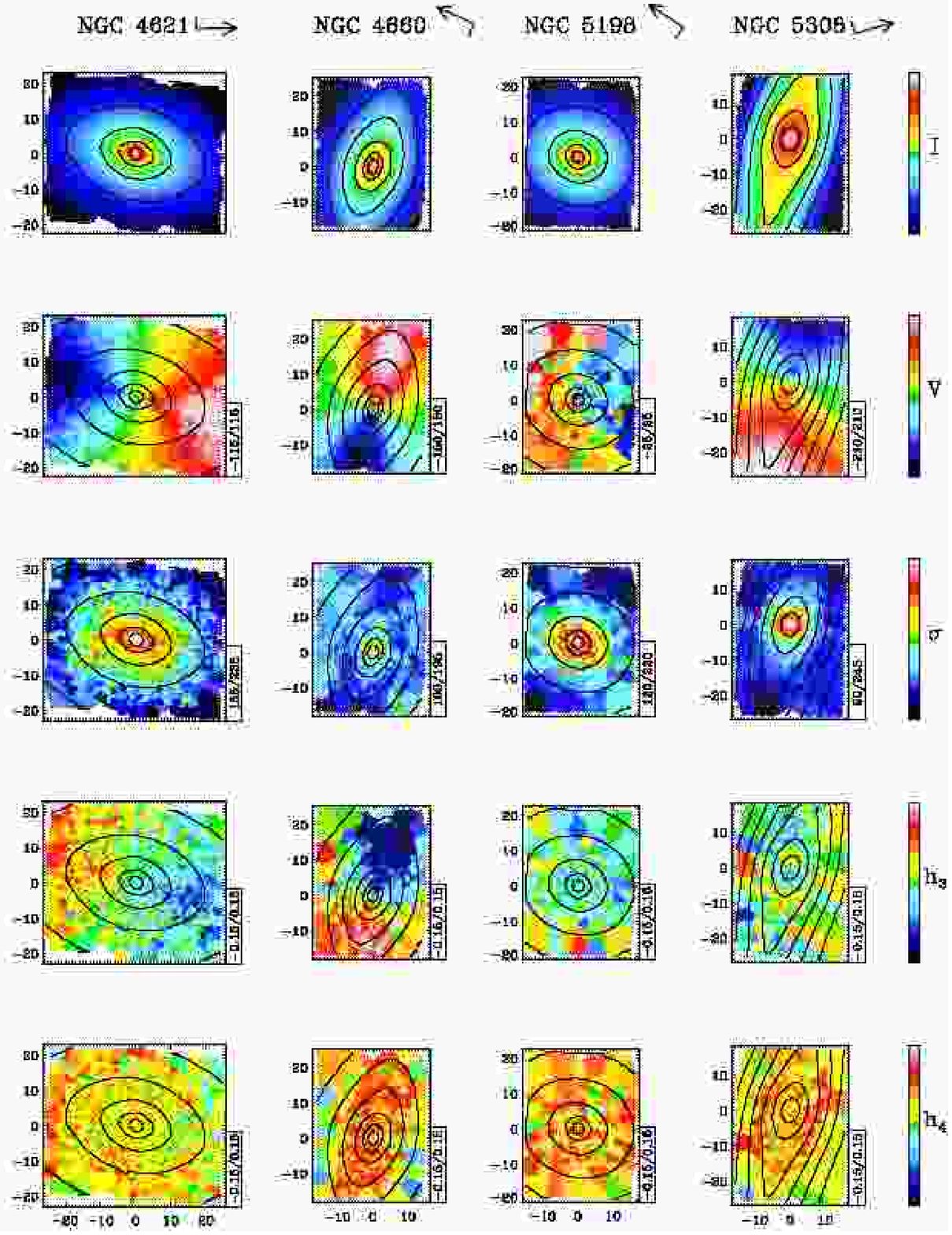}
\end{center}
\caption[]{}
\label{fig:maps10}
\end{figure*}

\addtocounter{figure}{-1}
\addtocounter{subfigure}{1}

\begin{figure*}
\begin{center}
  \includegraphics[width=\textwidth,trim=0cm 0cm 0cm 0cm]{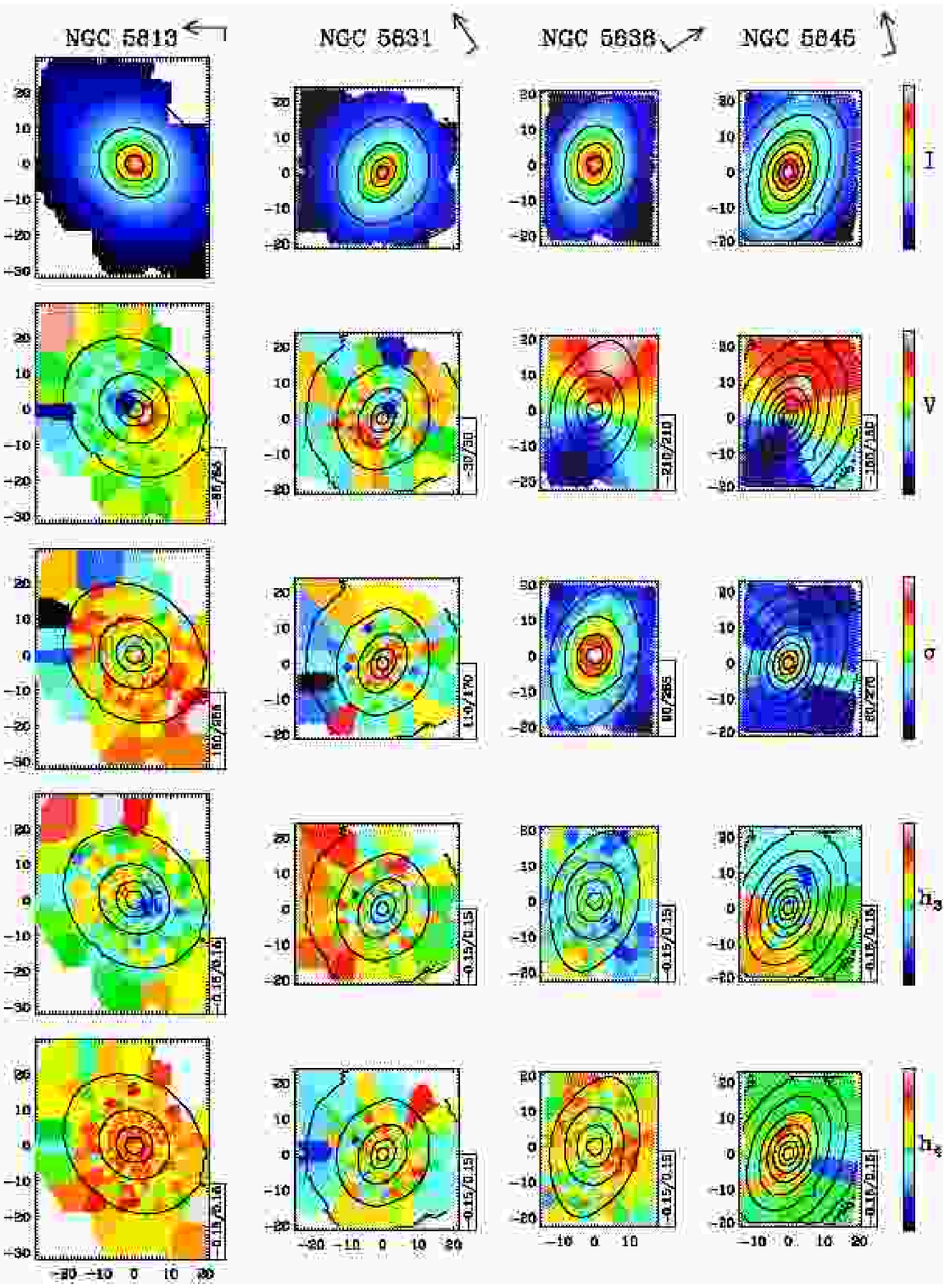}
\end{center}
\caption[]{}
\label{fig:maps11}
\end{figure*}

\addtocounter{figure}{-1}
\addtocounter{subfigure}{1}

\begin{figure*}
\begin{center}
  \includegraphics[width=\textwidth,trim=0cm 0cm 0cm 0cm]{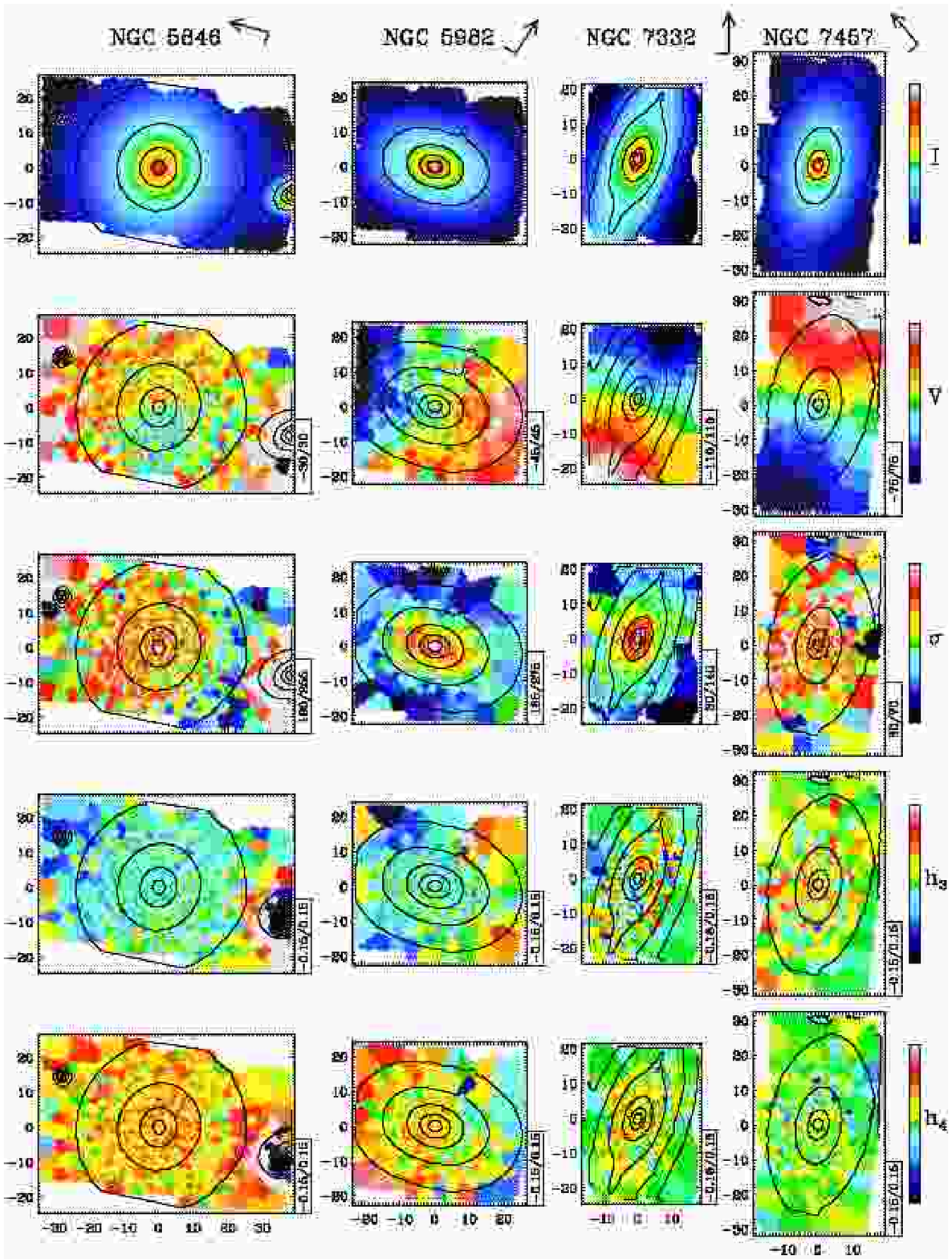}
\end{center}
\caption[]{}
\label{fig:maps12}
\end{figure*}

\section{Observed kinematics}
\label{sec:maps}

Figures~\ref{fig:maps1}-\ref{fig:maps12} display maps of the absorption-line kinematics of the 48
objects, all plotted on the same angular scale.
Not counting multiple exposures, and taking into account overlap of mosaiced exposures, these maps
correspond to a total of about 130000 galaxy spectra (about 32000 after adaptive binning). The maps
are displayed according to increasing NGC number. In each case we show the total intensity
reconstructed from the \sauron\ spectra, the mean stellar velocity $V$, the velocity dispersion
$\sigma$, as well as the Gauss--Hermite moments $h_3$ and $h_4$, as derived from the pPXF routine.
Mean velocities are with respect to estimated heliocentric systemic velocities,
the values of which are provided (corrected for the barycentric motion) in Table~\ref{tab:allgal}.

The maps display a wealth of structures. Most galaxies in this E/S0 sample show
a significant amount of rotation. Some galaxies exhibit isovelocity contours
steeply rising towards the photometric major-axis and also tend to show
elongated or dumbbell shaped $\sigma$ fields, for example NGC~821, NGC~2549,
NGC~3377, NGC~3384, NGC~4526, NGC~4660, NGC~5308 and NGC~5845. Other objects
have strongly misaligned photometric and kinematic axes, indicative of a
non-axisymmetric structure (e.g., a bar). This includes NGC~474, NGC~2699,
NGC~3384, NGC~4262 and NGC~4477. Some objects display an increase in $\sigma$
along the photometric major-axis associated with a flattening or turnover of
the mean velocity gradient along the same axis, for example NGC~4550 and
NGC~4473. Kinematically decoupled components (KDCs), showing either a twist
in the nuclear kinematic axis or a central velocity field clearly distinct
from the rest of the galaxy, are evident in a number of maps, for example
NGC~3414, NGC~3608, NGC~4458, NGC~5198, NGC~5813, NGC~5831, NGC~5982, and are
detected at smaller (apparent) scales in NGC~4150, NGC~4382, NGC~4621,
NGC~7332 and NGC~7457. Some of these were known previously, others are new.
Appendix~\ref{sec:list} contains a brief description
of these structures, including references to previous long-slit work on these
objects. The kinematics presented here will be analysed and discussed in
detail in subsequent papers of this series.

\section{Concluding remarks}
\label{sec:conclusions}

The maps presented in this paper are the result of the first survey of the stellar
kinematics of nearby early-type galaxies with an integral-field spectrograph.
This unique dataset demonstrates that early-type galaxies display significant and
varied structure in their kinematic properties, reinforcing the preliminary evidence shown in
Paper II, and the hints provided by earlier long-slit spectroscopy.
The present \sauron\ maps demonstrate that 2D coverage
is a prerequisite to properly understand the dynamics of nearby galaxies.

The maps shown here constitute only part of the information contained in the \sauron\ spectra. The
morphology and kinematics of the emission-line gas, as well as maps of the various line-strengths
derived from the same observations will be presented in future papers.

%
%
%
\section*{Acknowledgements}

We thank Kambiz Fathi, Jesus Falc\'on-Barroso, 
Marc Sarzi and Glenn van de Ven for their help and support at different stages of this work.
It is a pleasure to thank the Isaac Newton Group staff, in particular Rene Rutten,
Tom Gregory and Chris Benn, for enthusiastic and competent support on
La Palma. The \sauron\ project is made possible through grants
614.13.003 and 781.74.203 from ASTRON/NWO and financial contributions
from the Institut National des Sciences de l'Univers, the Universit\'e
Claude Bernard Lyon~I, the Universities of Durham and Leiden, the
British Council, PPARC grant `Extragalactic Astronomy \& Cosmology at
Durham 1998--2002', and the Netherlands Research School for Astronomy
NOVA.
This project made use of the LEDA database: http://leda.univ-lyon1.fr/.
MB acknowledges support from NASA through Hubble Fellowship grant
HST-HF-01136.01 awarded by Space Telescope Science Institute, which is
operated by the Association of Universities for Research in Astronomy,
Inc., for NASA, under contract NAS~5-26555. MC acknowledges support from a
VENI grant awarded by the Netherlands Organization for Scientific Research.
The Digitized Sky Surveys were produced at the Space Telescope Science Institute under U.S. Government grant NAG W-2166. The images of these surveys are based on photographic data obtained using the Oschin Schmidt Telescope on Palomar Mountain and the UK Schmidt Telescope.

%
%
%

%
%
\appendix
\section{Description for individual galaxies}

\label{sec:list}

Here we briefly comment on the structures observed
in the \sauron\ maps of the E/S0 sample presented in this paper.
A thorough and quantitative assessment of these structures will
be carried out in subsequent papers.
Independently derived velocity and velocity dispersion maps
for NGC~3379, NGC~3384, NGC~4550 and NGC~7457
were presented and discussed in \citet{Afa02}, \citet{Sil02} and \citet{Sil03}.

\begin{description} \item[\bf NGC 474:] This galaxy (Arp~227), famous for its shell structures
\citep[e.g.][]{Tur99}, is a clear case of a non-axisymmetric object, with a
strong misalignment between the kinematic and photometric major-axes
\citep{Hau96}.

\item[\bf NGC 524:] This galaxy has surprisingly high velocities (amplitude larger than 140~\kms)
considering its low ellipticity $\epsilon < 0.05$ \citep{Sil00, SP00}, as well as large $h_3$ values up to large
radii, anti-correlated with $V$.

\item[\bf NGC 821:] A close to edge-on galaxy with one of the most prominent disk-like kinematics in
the E/S0 \sauron\ sample revealed by its velocity field \citep*[see also][]{Pin03, SB95, BSG94}.

\item[\bf NGC 1023:] This is an example of an SB0 galaxy showing $h_3$ anti-correlated with $V$
inside $\sim 10\arcsec$, and correlated outside \citep{Bow01}. There is a strong twist of the zero velocity curve
near the center. Stellar kinematics has been previously published by \citet{SP97b}, \citet{Nei99}, \citet{Sil99},
including multiple long-slit data parallel to the major-axis \citep*{Deb02}.
The $h_3$ and $h_4$ maps show significantly negative values on the western side
of the galaxy which are not observed by \citet{Deb02}. The cause for this
discrepancy is not known, so these features should be confirmed.

\item[\bf NGC 2549:] The central high dispersion region is elongated along the photometric
minor-axis, due to the contribution of a rapidly rotating component
\citep[see also][]{Nei99, SP97b, Sei96}, exemplified by the
isovelocities pinched together along the photometric major-axis, and the correspondingly
large $h_3$ values.

\item[\bf NGC 2685:] The kinematics of this famous object \citep[the Helix galaxy;][]{Bur59,Pel93}
are strongly perturbed by extinction due to polar dust lanes on its North East side. The velocity
dispersion in the center displays a double peak structure \citep{SP97b, Hau99}.

\item[\bf NGC 2695:] This galaxy shows a high amplitude in $h_3$ in the central 7\arcsec,
anti-correlated with the mean velocity which has pinched isocontours. The
long-slit data of \citet{SP97a} contains a hint of this central kinematical structure.

\item[\bf NGC 2699:] The kinematical major-axis in the central region is misaligned from the
photometric major-axis. The mean velocity map shows a rapidly rotating component with a local
maximum at about 5\arcsec\ from the centre: within this region, $h_3$ departs significantly from
zero and is anticorrelated with $V$. There is a weak hint of such a decoupled structure
in the data of \citet{SP00}.

\item[\bf NGC 2768:] This galaxy shows a rather cylindrical velocity
field, with a shallow gradient across the rotation axis \citep{FI97}.
There is a pronounced dip in the central dispersion field, also clearly present in
the data of \citet{SP97b}. Significant dust extinction is present north of the center \citep{Mic99}.

\item[\bf NGC 2974:] A very rapidly rotating galaxy which has an $h_3$ field with one of the
strongest amplitudes. The $h_4$ map exhibits a central dip, as found by e.g., \citet{CvdM94},
but stays positive, in contrast to the data of \citet{BSG94}.

\item[\bf NGC 3032:] This dusty galaxy shows mild rotation and low dispersion values.

\item[\bf NGC 3156:] Another dusty galaxy with rather low dispersion everywhere ($\sigma <
75$~\kms), consistent with other authors \citep{RCF99, BSG94}.

\item[\bf NGC 3377:] The zero velocity curve is slightly misaligned with the photometric minor-axis
of the galaxy (Paper~I; \citeauthor{cop04} \citeyear{cop04}), giving the modest minor-axis rotation
observed in the long-slit study of \citet{Ha01}, and the dispersion map shows a clear
elongation along the photometric minor-axis of the galaxy.

\item[\bf NGC 3379:] This roundish galaxy has $h_3$ values being anti-correlated with the mean
velocity, and has fairly constant positive $h_4$ values: this is consistent with other published data
\citep[e.g., ][]{Ha01, G00, SS99}.

\item[\bf NGC 3384:] The dispersion map exhibits a dumbbell-like shape, coincident with a rapidly
rotating component in the central 5\arcsec\ \citep{F97, Bu96},
with a strong anticorrelation signature in $h_3$.
The velocities are nearly cylindrical in the outer part
of this galaxy classified as SB0 (Paper II). Our $h_4$ map shows a slight dip
at the centre, consistent with the long-slit profile of \citet{F97}. Our $h_4$ map
is consistent with $h_4$ being positive everywhere, in contrast to the measurements
of \citet{Pin03}.

\item[\bf NGC 3414:] A central counter-rotating component is visible in
the velocity map of this peculiar galaxy (Arp 162), with the $h_3$ field
showing a corresponding sign reversal just outside this kinematically
decoupled structure. The decoupled nature of this central component has
not been previously reported in the literature, although long-slit data exist
\citep{Ber95}.

\item[\bf NGC 3489:] The photometry is perturbed by dust extinction. The
effect is not evident in the kinematic maps although it may explain the local
increase in the dispersion map at about 15\arcsec\ West from the centre \citep{Ca00}.
Inside 4\arcsec\ the opening angle of the iso-velocities decreases,
indicating a fast rotating structure.
$h_3$ decreases inwards, with a turn over at a radius of about 4\arcsec,
inside of which it shows a small increase.

\item[\bf NGC 3608:] The velocity field displays the presence of the counter-rotating core
\citep{Jed88,Ha01} in the central 8\arcsec. As in the case of NGC~3384, we measure positive $h_4$
values everywhere, in disagreement with the data published by \citet{Pin03}.

\item[\bf NGC 4150:] The dispersion map exhibits a double peaked structure along the major-axis,
which corresponds to a region where the velocity gradient flattens out. A closer look at the
central 5\arcsec\ reveals the presence of a counter-rotating structure.

\item[\bf NGC 4262:] This strongly barred galaxy has a remarkably regular velocity field, the
long-axis of the bar being clearly misaligned with the kinematic minor-axis. $h_3$ changes sign
twice along the major-axis.

\item[\bf NGC 4270:] At large radii, the velocity field is close to cylindrical. Inside 10\arcsec\
the opening angle of the iso-velocity contours decreases, with a corresponding increase in the $h_3$
amplitude. The velocity curve in the central $\sim$5\arcsec\ is flat \citep{SP97a}.

\item[\bf NGC 4278:] This galaxy shows fairly regular kinematics, with the mean velocity decreasing
at radii larger than 10\arcsec\ along the major-axis \citep{SG79,DB88}.
The central dispersion peak is elongated along the major-axis.
The $h_4$ map reveals relatively high positive values everywhere except inside 6\arcsec\ where it
drops significantly, consistent with the long-slit data published by \citet{vdM93}, but
not with \citet{BSG94}.

\item[\bf NGC 4374:] The velocity map is consistent with $V\approx0$ everywhere in the \sauron\
	field \citep{D81}, except for a very weak large-scale gradient and a rotation pattern in the central
	3\arcsec. With NGC\,4278 and NGC\,4486, this is among the galaxies in the \sauron\ E/S0 sample
	with the highest relative contribution from emission lines.

\item[\bf NGC 4382:] The velocity field is nearly cylindrical in the outer part. It exhibits a
strong twist inwards, with the signature of a decoupled component in the central 2\arcsec, where
there is also a sign reversal in $h_3$ \citep{BSG94}. A clear counter-rotation is visible from higher resolution
{\tt OASIS} maps \citep{mcd04}.  The dispersion shows a remarkably strong depression in the central
8--10\arcsec, clearly seen in \citet{F97}. There is a ring of positive $h_4$ values at about
8\arcsec, hinted in the long-slit data of \citet{BSG94}.

\item[\bf NGC 4387:] This rather boxy galaxy \citep{Pel90} has a central dispersion drop, already
noticed by \citet{Ha01}.

\item[\bf NGC 4458:] The KDC is clearly visible in the \sauron\ velocity map of this galaxy
\citep[see][]{Ha01}, and the velocity outside 5\arcsec\ is consistent with being equal to zero.

\item[\bf NGC 4459:] The small opening angle of the iso-velocities in the central 10\arcsec\
corresponds to a region where $h_3$ is anti-correlated with $V$, and has a rather high amplitude
\citep[see][]{Pet78}.

\item[\bf NGC 4473:] The velocity dispersion field exhibits a complex morphology, with a region of
high dispersion along the major-axis which widens at larger radii, and a central drop inside
3\arcsec, not observed in \citet{BSG94}.
The widening in the dispersion map is particularly significant outside 10\arcsec\ where
the mean velocity is observed to decrease outwards. This also corresponds to a region of lower
$h_4$.

\item[\bf NGC 4477:] The \sauron\ field of view zooms in on the bar of this
galaxy where the kinematic and photometric major-axis are misaligned
and the velocity field shows a significant twist of its zero velocity
curve within the central 4\arcsec\, and nearly cylindrical rotation
outside. \citet{Jar88} noticed the existence of considerable minor-axis 
rotation as well as a flat velocity dispersion profile with a central dip
which is not visible in our data.

\item[\bf NGC 4486:] This well-known giant elliptical galaxy has velocities consistent with zero
everywhere in the \sauron\ field of view, confirming the absence of rotation, and high central velocity
dispersion \citep{Sar78,DB88,BSG94,vdM94}.

\item[\bf NGC 4526:] This S0 object has a prominent dust disc which is not only visible in the
reconstructed \sauron\ image, but also influences the observed kinematics. The dumbbell shape in the
dispersion map is due to a fast rotating stellar component, well visible in the velocity map.
The data of \citet{Pel97} also shows
the drop in velocity between 10\arcsec and 20\arcsec, which corresponds to this
component. $h_3$ reverses sign at about 20\arcsec\ along the major-axis.

\item[\bf NGC 4546:] The velocity field of this barred galaxy displays a regular rotation
pattern \citep{Bett97}, anticorrelated with $h_3$.

\item[\bf NGC 4550:] This object has two counter-rotating stellar disks of similar mass
(\citeauthor{Rix92}~\citeyear{Rix92}) which, at the spectral resolution of \sauron, produce a large
region with a near zero mean velocity $V$ and a dispersion increasing outwards along the major-axis.
Note the extraordinary decoupling revealed by \sauron, the outer disk being counter-rotating with
respect to the main body of the galaxy further away from the equatorial plane, also noticed by
\citet{Afa02}. $h_4$ is significantly lower outside 5\arcsec\ along the major-axis.

\item[\bf NGC 4552:] We detect weak rotation in this galaxy and a corresponding weak correlated
$h_3$ field \citep{BSG94}. The zero velocity curve is significantly twisted.

\item[\bf NGC 4564:] This flattened galaxy shows the signature of a disk-like component in the
pinched $V$ map, anti-correlated with $h_3$ \citep{BSG94, Ha01}.
The large $h_3$ amplitude along
the major-axis outside $\sim 5\arcsec$ is however not observed by \citet{Pin03} who
also measured significantly negative $h_4$.

\item[\bf NGC 4570:] The inner part of the velocity field and $h_3$ map suggest the presence of
a fast rotating disk-like structure. This component was also seen in long-slit observations
\citep{vdb98} and interpreted as due to bar-driven evolution \citep{vdbe98}.

\item[\bf NGC 4621:] This object is another case showing indication for a disk-like component
with strong anticorrelation between $V$ and $h_3$ \citep{b90,BSG94}. A
counter-rotating component is also detected inside 2\arcsec, previously discovered with adaptive
optics assisted integral-field spectroscopy \citep{W03}.

\item[\bf NGC 4660:] The velocity map shows the superposition of two rapidly rotating components: an
inner one inside 5\arcsec\ and a thicker structure outside 7\arcsec.
This galaxy was used by \citet{BSG94} as a prototype for discy ellipticals, 
due to its pointed isophotal shape and the strong $h_3$ anticorrelated with $V$.

\item[\bf NGC 5198:] The velocity field reveals a central KDC, rotating nearly perpendicularly with
respect to the outer body, both being misaligned with respect to the photometric axes of the galaxy.
A ``suspect'' central velocity profile was noticed by \citet{Jed89}.

\item[\bf NGC 5308:] This is a typical case of a close to edge-on disk galaxy showing a rapidly
rotating component \citep{Sei96} and a double sign reversal in $h_3$. The
central dispersion structure is box-shaped.

\item[\bf NGC 5813:] A galaxy with a well-known KDC \citep{Ef80, Ef82, Kor84, BSG94} evident in the
\sauron\ velocity map (see Paper~II). The galaxy has negligible rotation
at larger radii, and shows weak evidence for minor-axis rotation.
The dispersion field also shows an unusual ring-like
depression at about 5\arcsec\ from the center.

\item[\bf NGC 5831:] The \sauron\ maps needed significant rebinning to reach the
required $S/N$, but the velocity field reveals the well-known KDC \citep{Dav83, Pel90}.

\item[\bf NGC 5838:] This boxy galaxy is a strong rotator, with a double sign reversal in $h_3$.
Minor-axis spectra of \citet{Fal03} show a small amount of rotation,
very probably due to the extinction produced by a dust lane most prominent
on the South-West side within 3\arcsec\ of the centre.

\item[\bf NGC 5845:] The velocity field of this galaxy shows two distinct peaks on each side, at
about 2\arcsec\ and 8\arcsec\ from the center. These features are barely detected in the long-slit
data of \citet{SP02}. The central kinematically decoupled structure
corresponds to a lower dispersion and a peak in $h_3$.

\item[\bf NGC 5846:] This is an example of a giant elliptical with a very low (but present) rotation
\citep{Car93},
and a high central dispersion. The \sauron\ field of view includes a foreground star and a companion
(NGC~5846A) North and South of its nucleus, respectively. $h_4$ is high everywhere, confirming
the trend observed by \citet{BSG94}.

\item[\bf NGC 5982:] The \sauron\ data confirm the presence of a KDC first
detected by \citet{Wag88}. The \sauron\ maps reveal that
the central rotation axis seems to be nearly
perpendicular to the rotation of the main body, the central $h_3$ structure
following this orientation.

\item[\bf NGC 7332:] Another strongly boxy galaxy with a KDC (counter-rotating) in the central
3\arcsec\ \citep[see][]{Fal04}, and a high central dispersion region elongated along the major-axis
\citep[see also data by][]{FIF94, Sei96, SP97b, Sil99}.

\item[\bf NGC 7457:] The velocity map exhibits a nearly cylindrical rotation, as well as a central
counter-rotating structure \citep[see also data by][]{SP97b}.
\end{description}

\section{Tests of the kinematics extraction}

\subsection{Comparison of pPXF with FCQ}
\label{sec:fcq}

\begin{figure}
\begin{center}
  \includegraphics[width=\columnwidth]{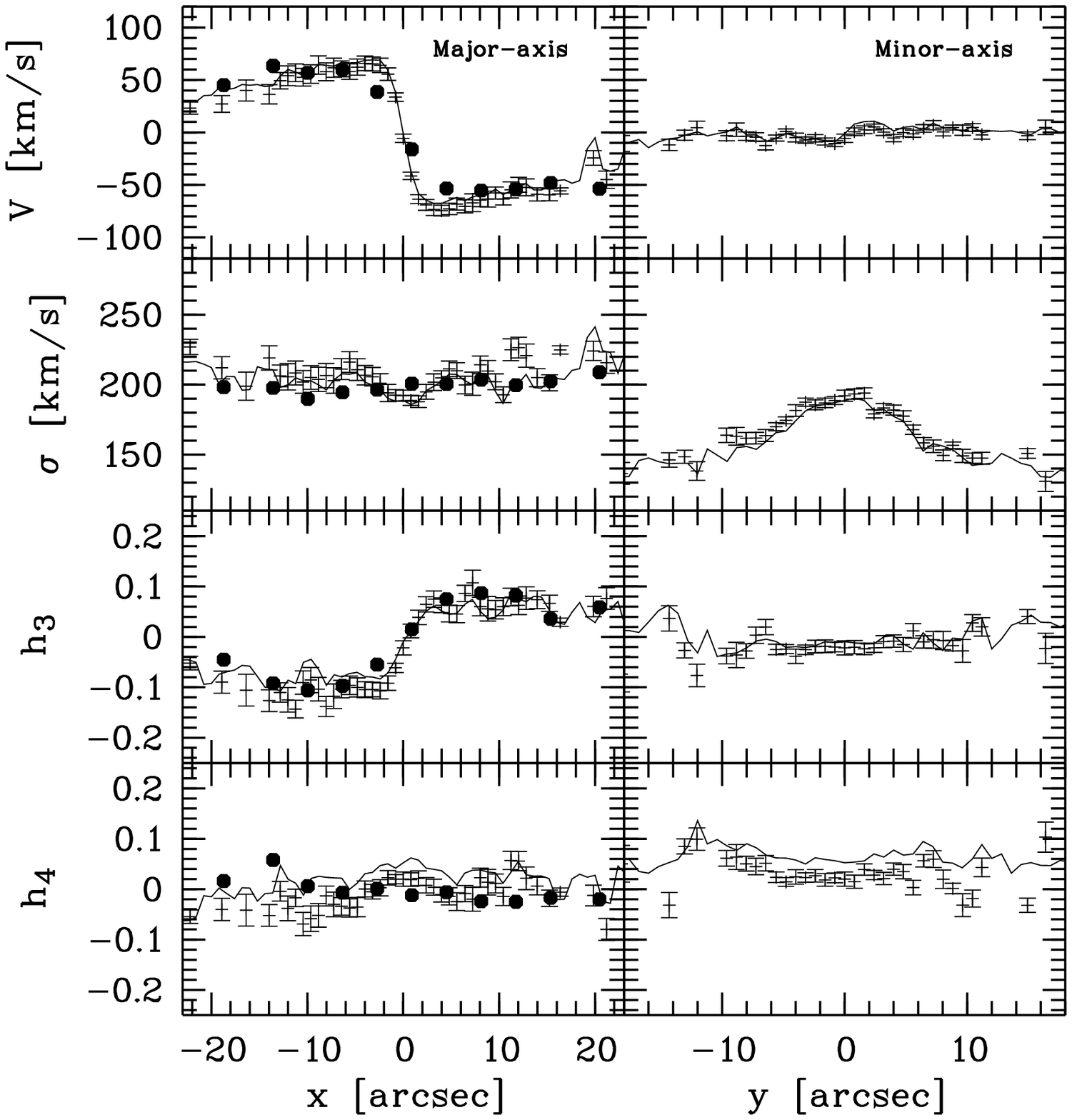}
\end{center}
\caption[]{Comparison between the stellar kinematics of NGC~4473 extracted via the 
FCQ (crosses) and the pPXF (solid lines) methods. 
The major and minor-axis profiles are respectively shown in the left and
right panels, with, from top to bottom $V$, $\sigma$, $h_3$ and $h_4$. The
long-slit major-axis measurements of \citet{BSG94} are shown (filled dots) for
comparison.}
\label{fig:fcq}
\end{figure}

In Fig.~\ref{fig:fcq} we show a comparison between the stellar kinematics recovered
via FCQ and our pPXF technique (see Sect.~\ref{sec:pxf}).
We  illustrate this with the case of NGC~4473 which is free of (detected) emission lines
(see Fig.~\ref{fig:optimal}). In Fig.~\ref{fig:fcq} we present the major (left panels) and minor-axis
(right panels) profiles for the first four extracted parameters, namely $V$, $\sigma$, $h_3$ and
$h_4$. The agreement between the FCQ (crosses with error bars) and pPXF (solid lines) methods
is excellent except for a slight discrepancy observed in the $h_4$ profiles. FCQ tends
to provide $h_4$ values systematically lower than pPXF. Although this offset depends
significantly on the continuum subtraction procedure (e.g., degree of
polynomial) and the spectral domain (including the region around
H${\beta}$ or not) used in FCQ, we could not fully reconcile the FCQ and pPXF $h_4$ measurements.
This difference in the $h_4$ measurements may be related to the 
negative bias in the determination of $h_4$ via FCQ, when the LOSVD is not
well sampled, noticed by \citet{Jos01} (see their Fig.~13) and in \citet{Pin03}.

\subsection{Examples of resulting fits}
\label{sec:fits}

\begin{figure}
\begin{center}
  \includegraphics[width=\columnwidth]{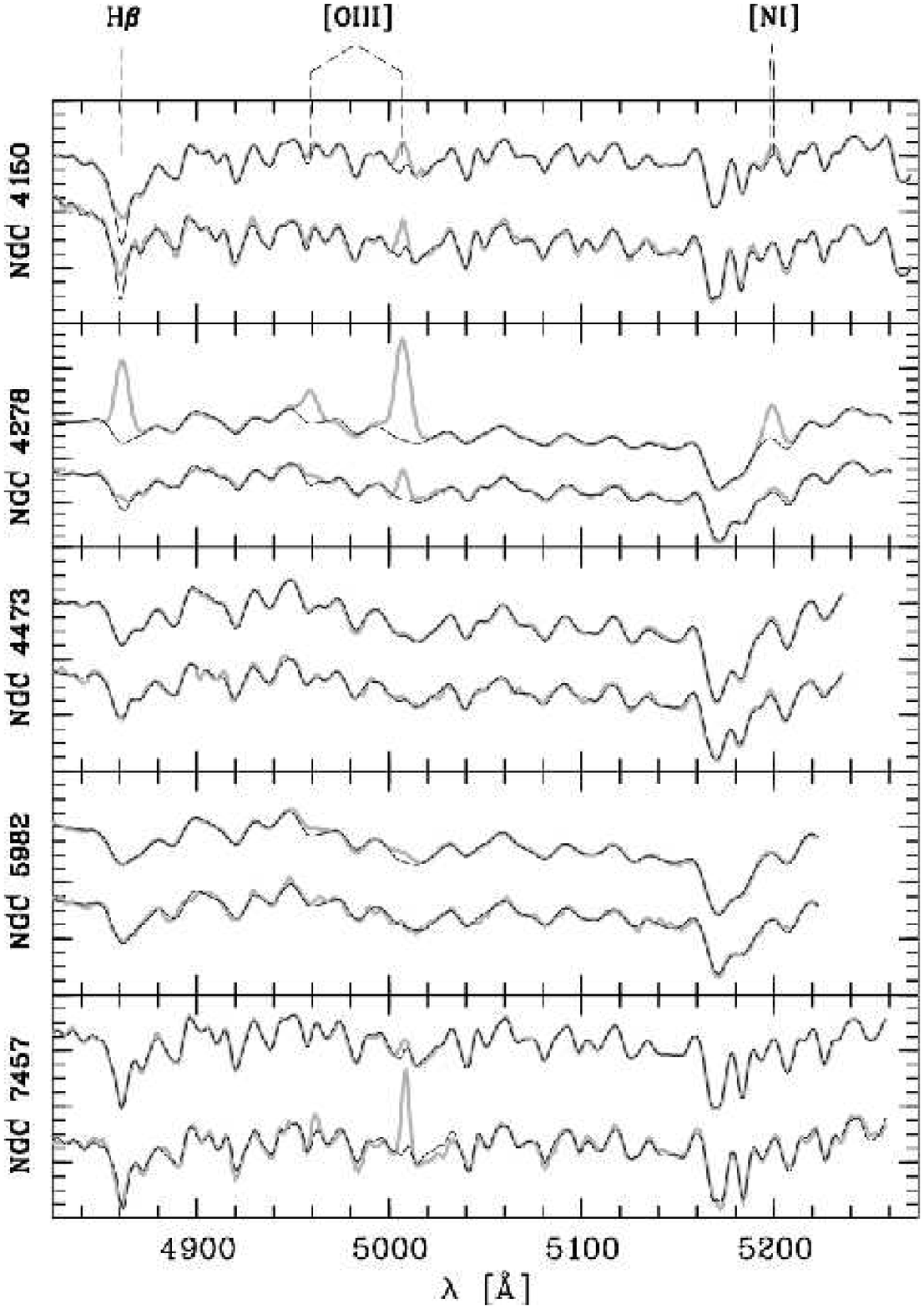}
\end{center}
\caption[]{Resulting spectral fits obtained with the pPXF routine for a few galaxies in the \sauron\
E/S0 sample. Each panel corresponds to a different galaxy (see name label at the left),
with the upper spectrum taken at the center and the lower one at a radius
of about 10\arcsec. The \sauron\ spectra are drawn as thick grey lines, and the
corresponding fit obtained from the pPXF routine drawn as thin solid lines. All spectra 
have been shifted to rest wavelengths, as well as normalized and vertically offset for better legibility.
The expected locations of a few emission lines are indicated at the top of the Figure.}
\label{fig:optimal}
\end{figure}

Here we present a few illustrative examples of spectral fits
obtained using the pPXF and optimal fitting procedure described in Sect.~\ref{sec:kinematics}.
Figure~\ref{fig:optimal} displays the results obtained for five galaxies
in the \sauron\ E/S0 sample, probing a range in absolute luminosity,
absorption-line depths and degree of contamination by emission lines:
\begin{itemize}
\item NGC~4150 which has rather deep H$\beta$ absorption lines
and some contribution from emission lines,
\item NGC~4278 which shows very significant emission line contamination,
\item NGC~4473, an intermediate luminosity elliptical with no detected emission lines,
\item NGC~5982 as the brightest galaxy in the \sauron\ E/S0 sample, and
\item NGC~7457 for a case of a low dispersion galaxy with detected emission.
\end{itemize}

In each case, we present a central spectrum as well as a spectrum at a radius of about 10\arcsec\
(chosen to lie at 45\degr from the photometric major-axis).
As mentioned in Sect.~\ref{sec:kinematics}, all fits
are excellent, with local discrepancies between the fitted and observed spectra
emphasizing the presence of emission lines (H$\beta$, [{\sc O$\,$iii}], [{\sc N$\,$i}]). The use of an individual
optimal template spectrum for each \sauron\ spectrum is crucial as line depths can vary significantly
within the \sauron\ field of view. These plots demonstrate the flexibility of our procedure.

\subsection{Template mismatch}
\label{sec:mismatch}

The effect of template mismatch is difficult to quantify, as galaxy spectra are a complex
(luminosity-weighted) mix of different stellar populations with different velocity distributions.
We can however perform some simple tests by examining the change in the measured stellar kinematics
as we perturb the spectral shape of either the stellar template or the galaxy spectrum.

We performed this in a way similar to the procedure described by \citet{RW92}. We first chose an optimal template representative of the ones obtained for the galaxies in the E/S0 \sauron\ sample. We added a random Gaussian perturbation to that template {\em before} convolving it with a LOSVD parametrized as a Gauss-Hermite series to simulate a `galaxy' spectrum. In contrast to what is done in \citet{RW92}, we preferred not to add any additional Poissonian noise, to isolate the sole effect of template mismatch. We then retrieved the kinematic parameters with our pPXF routine (see Sect.~\ref{sec:pxf}) to finally compare the extracted and input values. This is repeated for many different Monte Carlo realizations of the perturbation.

\begin{figure}
\begin{center}
  \includegraphics[width=\columnwidth]{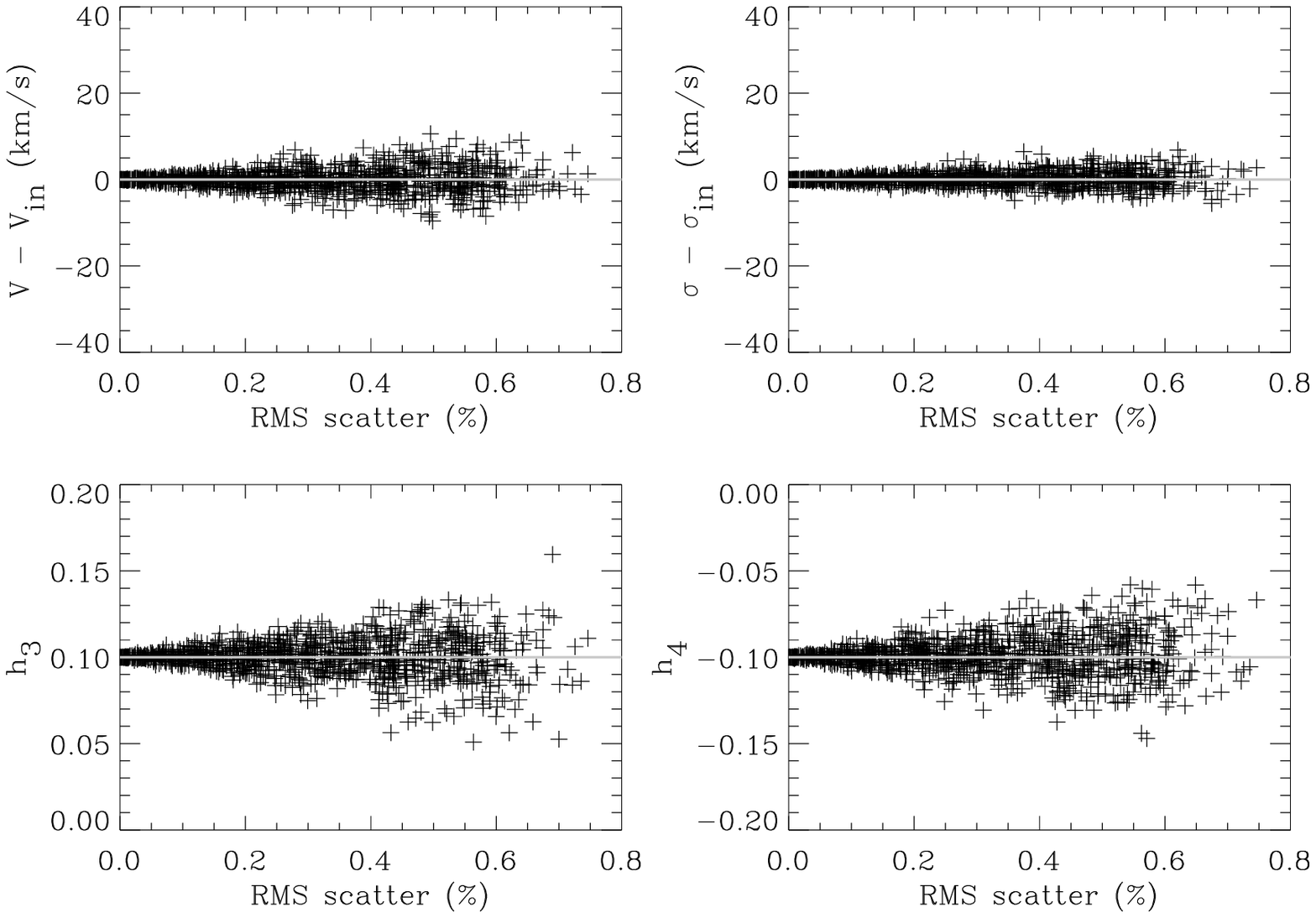}
\end{center}
\caption{Effect of template mismatch on retrieved stellar kinematics. A random perturbation was added to a template spectrum {\em before} convolving it with a LOSVD parametrized as a fourth order Gauss-Hermite series. The input LOSVD parameters were $V_{\rm in} = 20$~\kms, $\sigma_{\rm in} = 200$~\kms, $h_{3,{\rm in}} = 0.1$ and $h_{4,{\rm in}}= -0.1$. The measured kinematics (crosses) for 1000 different Monte Carlo realizations of the perturbation, are shown as a function of the corresponding RMS (in percentage) of the residual mismatch between the input spectrum and the best fitting model (see text). The thick gray line indicate the true input parameters of the LOSVD.}
\label{fig:mismatch}
\end{figure}

In Figure~\ref{fig:mismatch}, we show such a comparison for an input velocity of
$V_{\rm in} = 20$~\kms, a dispersion of $\sigma_{\rm in} = 200$~\kms,
$h_{3,{\rm in}} = 0.1$ and $h_{4,{\rm in}}= -0.1$. The plots show the retrieved
kinematics versus the RMS of the residual mismatch between the simulated
`galaxy' spectrum and the best fitting model. The maximum uncertainty in the
retrieved kinematic parameters increases almost linearly with the scatter in
the fit. In the case of the \sauron\ data presented in this paper, we
measured (on the highest $S/N$ spectra where template mismatch dominates over
Poissonian noise), a typical residual mismatch in the fit of about
0.4\% RMS , with a maximum value of 0.55\%. The above test then indicates an
upper limit to any systematic error on $h_3$ and $h_4$ of $\la 0.03$,
although in practice our actual error is likely to be significantly lower
than this maximum.

\label{lastpage}
\end{document}